\documentclass[a4paper,10pt]{article} %%% option by default two columns
\usepackage[textwidth=16.5cm,textheight=22cm]{geometry}
\usepackage{amsmath}
\usepackage{graphicx}
\usepackage{epstopdf}
\usepackage[numbers]{natbib}% numbered references
\usepackage{hyperref}
\usepackage{url}
\usepackage{CJKutf8}

\hypersetup{colorlinks=true,citecolor=blue,urlcolor=blue,linkcolor=blue}
\usepackage{ragged2e}
\usepackage[ruled]{algorithm2e}
\usepackage{amssymb}
\usepackage{import}
\usepackage{tikz}
\usepackage{pgfplots}
\usepackage{xcolor}
\usepackage{multirow}
\usepackage{makecell}
\usepackage{placeins}

\begin{document}

\definecolor{spec_blue}{rgb}{0.37, 0.55, 0.82}
\definecolor{spec_green}{rgb}{0.37, 0.82, 0.37}
\definecolor{spec_brown}{rgb}{0.62, 0.35, 0.17}
\definecolor{spec_amethyst}{rgb}{0.6, 0.4, 0.8}
\definecolor{spec_orange}{rgb}{1.0, 0.49, 0.0}

\renewcommand{\labelenumii}{\arabic{enumi}.\arabic{enumii}}

\begin{center}
    \Large{\textbf{Building Acoustics 01: Finite Element Model of an Building Acoustics Test Facility to Predict the Sound Transmission Loss Based on DIN EN ISO 10140}} \\
\end{center}
\begin{center}
    Sebastian Schmidt$^1$, Sabine C. Langer$^1$ \\
    \vspace{0.5cm}
    $^1$ Institute for Acoustics and Dynamics, Technische Universität Braunschweig, Langer Kamp 19, 38106 Braunschweig, Germany \\
    \vspace{0.5cm}
    18.05.2026
\end{center}

\begin{center}
    \begin{minipage}{0.8\textwidth}
        \textbf{Abstract} - In the context of building acoustics, sound transmission loss estimations are crucial to quantify the noise pollution in buildings. When developing building prototypes in the sense of an acoustic-oriented design process, it is desirable to have an virtual prototype, especially in early development stages, to estimate, for instance, the influence of different material or geometry configurations on to the sound transmission loss. This contribution aims to present a simple virtual prototype of an building acoustics test facility in accordance with DIN EN ISO 10140 for the measurement of the sound transmission loss of single- and double-leaf walls with and without insulation. Here, the finite element method is used as the numerical modelling method of choice. In the course of this, geometry and mesh creation was done using SALOME 9.14 whereas the institute's in-house research code elPaSo was utilised for the matrix assembly and solving procedure. At first, elPaSo was verified by the commercial software COMSOL 6.3 considering a small-scale test facility. Afterwards, the large-scale test facility finite element model was created using a frequency- and domain-specific discretisation approach. The sound transmission loss of three different test specimens was estimated in one-third-octave bands from $8\,\text{Hz}$ to $630\,\text{Hz}$, where the double-leaf wall with insulation exhibited good agreement to the theoretical sound transmission loss profile from literature. 
    \end{minipage}
\end{center}
\vspace{1cm}

\section{Introduction}
Negative effects of noise on human health are well known which is why the reduction of noise in buildings is an important topic. It is therefore necessary to incorporate noise reduction measures during the design process of buildings in the sense of an acoustic-oriented design process \cite{LAN2025}. Regulations and laws regarding noise protection in buildings are manifested in DIN EN ISO 4109-1 \cite{DIN4109-1}. One important quantity to estimate noise protection in buildings is the sound transmission loss (STL), describing the sound transmission through walls.\\
The STL can be measured in-situ in buildings or detached from the building in building acoustics test facilities fulfilling the regulations defined in DIN EN ISO 10140-5 \cite{DIN10140-5}. To gain a unified STL measure of different wall constructions, the measurement in test facilities is used instead of in-situ measurements. Nonetheless, on the one hand, the experimental setup is costly and measurements from different test facilities exhibit significant variability \cite{MEI1999}\cite{SCH1999}. On the other hand, especially in early design phases of an acoustic-oriented design process, material parameters and construction details of walls may not be fully specified. As a consequence, it is desirable to create a virtual prototype of a building acoustics test facility to  predict the STL numerically without relying on repeated time and resource consuming real-life experiments. \\
Under various numerical modelling approaches for sound pressure field predictions in rooms, this contribution focusses on the finite element method (FEM) due to its wave-resolving property and the associated advantageous spatial discretisation of the sound pressure field. Nevertheless, it should be mentioned that this advantage comes with a challenge when modelling large geometries and considering a wide frequency range as it requires then vast computational resources to compute the sound pressure field. The reason is the sound waves' decreasing wavelength with increasing frequency. To guarantee a good resolution of the different sound wave phenomena, at least six or four finite elements per wavelength are required using linear or quadratic shape functions \cite{ATA2015}. As a consequence, very fine discretised FE meshes arise with many degrees of freedom, resulting in a large linear system of equations. Using other established methods instead, e.g. the statistical energy analysis (SEA) \cite{ATA2015}\cite{LYO1994}, might be a viable option. Yet, statistical methods are not wave-resolving; thus, they do not provide spatial information about the sound fields, which is especially important when investigating complex constructions. \\
In the context of predicting sound fields in rooms, FEM is already used since 50 years; a comprehensive literature review is given in \cite{PRI2023}. The following brief literature review focusses on FE models related to building acoustics test facilities. One may start with Craggs and Stead \cite{CRA1976} who modelled the sound transmission through a plate connecting two fluid cavities - however, the dimension of the problem was reduced due to numerical effort. Maluski and Gibbs \cite{MAL2000} investigated the sound transmission through a single-leaf wall in low-frequency ranges. At first, a geometrically downscaled model ($\text{room volumes}\,\leq1\,\text{m}^3$) was used for validation in the frequency range $1$ to $800\,\text{Hz}$, then a full-scale model ($\text{room volumes:}\,35\,\text{m}^3\,\text{and}\,40\,\text{m}^3$) was evaluated in the frequency range $31.5$ to $160\,\text{Hz}$. Ackermann \cite{ACK2002} computed the STL of different wall constructions. That said, the numerical example that most closely corresponds to a test facility as discussed in this contribution measured the STL of a single- and double-leaf wall considering room volumes $\leq1.5\,\text{m}^3$ in the frequency range $1$ to $1000\,\text{Hz}$. Furthermore, Langer and Antes \cite{LAN2003} coupled FEM with the Boundary Element Method (BEM) to estimate the STL of windows in the frequency range $100$ to $800\,\text{Hz}$ considering a room volume $\leq1\,\text{m}^3$. Next to this, Clasen \cite{CLA2008} applied scaling rules to downscale a DIN-conforming building acoustics test facility (factor $1$:$10$) and computed the STL as well as flanking transmissions in the context of single- and double-leaf walls up to $3200\,\text{Hz}$. Also, Wulkau \cite{WUL2011} used a downscaled test facility (factor $1$:$10$) to numerically estimate the STL. Even if the use of downscaled models seems to be a feasible option at this point, Kling \cite{KLI2008} pointed out the fact that original materials cannot be used in downscaled models due to frequency-dependent properties. Thus, applicable materials need to be found, which behave similar to the original materials in the scaled frequency range. Lastly, the authors Arjunan et al. \cite{ARJ2014} set up a finite element model of a full-scale building acoustics test facility with respect to ISO 10140. The model was used to predict the STL of stud based double-leaf walls in one-third octave bands ($100$-$3150\,\text{Hz}$). Therefore, the sound pressure fields were computed at 48 frequencies, namely each central frequency of the one-third octave bands and $+5\,\%$, $-5\,\%$ of the respective central frequency. \\
At this point, it can be concluded that several promising investigations are undertaken towards the STL prediction using a numerical model of an building acoustics test facility. Nonetheless, it should also be mentioned that the investigations are limited to small geometries or downscaled test facilities, and, when considering a large-scale test facility, small frequency ranges or a small number of frequency evaluation points. \\
This article aims to present an open-source dataset of a simple full-scale building acoustics test facility finite element model considering the regulations of DIN EN ISO 10140 to estimate the STL of a single- and double-leaf wall with and without insulation in one-third octave bands from $8$ to $630\,\text{Hz}$. The corresponding dataset has recently been issued as a benchmark \cite{SCH2025} in the EAA Computational Acoustics community on Zenodo \cite{EAARef}. In doing so, the model is created and meshed using SALOME 9.14 while the matrices are assembled and solved using the institute's in-house research code elPaSo (elementary Parallel Solver) written in C\texttt{++} \cite{SRE2023}, whose efficiency has already been proven, for instance, in \cite{BLE2024} and \cite{SRE2021}. 
The development and deployment of elPaso follow the SURESOFT principles for sustainable research software development, emphasizing reproducible results and the software’s capacity to endure and evolve over time \cite{BLE2022}. \\
The article is structured as follows. Section \ref{Sec:BATF} summarises the key aspects of the building acoustics test facility with respect to DIN EN ISO 10140 that will be covered by the FE model. Then, Section \ref{Sec:VMFEM} introduces the FE formulations of the test facility's subdomains as well as the fluid-structure coupling. Following this, Section \ref{Sec:BATF_FEM} gives a brief summary of the test facility FE model derived from the key aspects in Section \ref{Sec:BATF}. The simulation chain including SALOME and elPaSo is verified using the commercial FEM software COMSOL 6.3 by considering a small-scale building acoustics test facility FE model in Section \ref{Sec:elPaSoVer}. Lastly, the results of the numerical STL estimation utilising the large-scale building acoustics test facility FE model are presented in Section \ref{Sec:Res_elPaSo_LSBATF}.

\section{Building Acoustics Test Facility with respect to DIN EN ISO 10140}\label{Sec:BATF}
\begin{figure}[!b]
    \centering
    \def\svgwidth{\linewidth}
    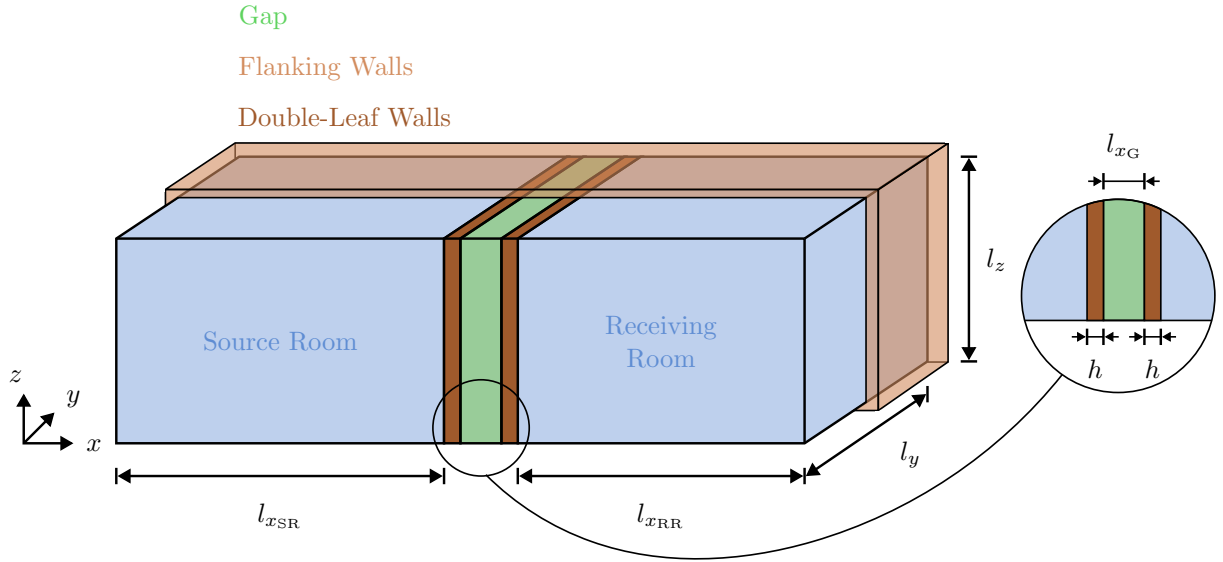
    \caption{Schematic sketch of building acoustics test facility consisting of a source and receiving room as well as a double-leaf wall as a test specimen. For illustrative purposes, only half of the flanking walls are shown. The measures $l_i$ define the dimensions in the respective coordinate directions $i=x,y,z$, while the subscripts SR, RR and G specify the dimensions of the source room, receiving room and gap. $h$ is the test specimen's wall thickness.}
    \label{fig:BATF}
\end{figure}

Building acoustics test facilities are used to estimate the airborne sound insulation as well as the footfall sound insulation. Here, the focus is laid on estimating the airborne sound insulation, which means, the STL. Figure \ref{fig:BATF} shows a schematic sketch of an building acoustics test facility, consisting of two rooms - the source and receiving room - as well as, depending on the test specimen, two walls separated by a gap that can be filled with insulation (double-leaf wall with and without insulation) or a single wall (single-leaf wall). 
 \\
As the name suggests, the source room is exited by sound sources whose sound is radiating into the receiving room due to the following vibroacoustic chain. The sound sources in the source room evoke airborne sound which excites the adjacent walls to structure-borne oscillations. Depending on the test specimen's configuration, the oscillating wall adjacent to the gap also excites the gap-filling air or insulation to oscillations. If no insulation is applied, pure airborne sound propagates within the gap; and if a porous material is used as insulation, airborne and structure-borne sound propagate within the porous material that consists of a solid and fluid phase, where the solid phase is also called frame. Again, the airborne (and structure-borne) sound acts as a load onto the receiving room's wall adjacent to the gap. Consequently, the wall radiates airborne sound into the receiving room. Since flanking transmission is neglected in the benchmark model presented here, the vibroacoustic chain is not extended to the flanking walls. \\
A convenient approach to measure the airborne sound insulation of the test specimen is to compare the sound pressure levels in the source and receiving room. This is done by estimating the STL $R$ defined in accordance with DIN EN ISO 10140-2 \cite{DIN10140-2} as
\begin{equation}\label{eqn:STL}
    R = L_1-L_2 + 10\text{lg}\frac{S}{A},
\end{equation}
where $L_1$, $L_2$ are the energy-averaged sound pressure levels in the source and receiving room, $S$ is the area of the test specimen, and $A$ the equivalent absorption area in the receiving room. DIN EN ISO 10140-4 \cite{DIN10140-4} defines the energy-averaged sound pressure levels $L_i$, $i=1,2$, as
\begin{equation}\label{eqn:EnerAveSP}
    L_i = 10\text{lg}\frac{p_{i,1}^2+p_{i,2}^2+\cdots+p_{i,d}^2}{d\,p_0^2},
\end{equation}
wherein $p_{i,j}$, $j=1,...,d$, is the root-mean-square-value of the sound pressure measured at $d$ positions in the source and receiving room, and $p_0=2e{-5}\,\text{Pa}$ is the reference sound pressure. According to DIN EN ISO 10140-2 \cite{DIN10140-2}, it is common to measure the root-mean-square-value of the sound pressure in one-third octave bands with mid-frequencies $100\,\text{Hz}$ to $5000\,\text{Hz}$. Hence, the root-mean-square value of the sound pressure measured at a position $j$ in one-third octave bands with the lower, upper and mid-frequency $f_l$, $f_u$ and $f_m$ is calculated as \cite{VIG2008}
\begin{equation}
    p_{i,j}(f_m) = \frac{1}{\sqrt{2}}\left( \sum_{f=f_l}^{f_u} |\hat{p}_{i,j}(f)|^2 \right)^\frac{1}{2},
\end{equation}
where $|\cdot|$ denotes the absolute value and $\hat{p}_{i,j}\in\mathbb{C}$ is the complex measured sound pressure at position $j$ in room $i$ and frequency $f$. Subsequently, complex valued variables are assigned with a $\hat{\Box}$. The lower, upper and mid-frequency $f_l$, $f_u$ and $f_m$ are calculated based on DIN EN 61260-1 \cite{DIN61260-1}. As a consequence of defining the sound pressure in one-third octave bands, the STL $R$ is then also defined in one-third octave bands.\\
\\
To ensure reproducible STL measurements, building acoustic test facilities as well as the STL measurement procedures are standardized by DIN EN ISO 10140. In the following, the key aspects covered by the FE Model in Section \ref{Sec:BATF_FEM} will be briefly reviewed.\\
DIN EN ISO 10140-5 \cite{DIN10140-5} defines the requirements regarding the test facility itself; beginning with the fact that the test facility consists of two adjacent rooms between which the test specimen is placed. The rooms' minimal volume is $V=50\,\text{m}^3$, whereat both rooms should have a minimal volume difference of $10\,\%$. Table \ref{tab:DimBATF} lists the measures of the contribution's building acoustics test facility that meet the DIN-requirements listed above. 
\begin{table*}[!b]
    \label{tab:DimBATF}
    \caption{Dimensions of the building acoustics test facility's source and receiving room.}\label{tab:DimBATF}
    \vspace{0.3cm}
    \centering
        \begin{tabular}{lcccc}
        \hline
                        & $l_x$ in m & $l_y$ in m & $l_z$ in m &  $V$ in $\text{m}^3$ \\
        \hline
        Source Room     & $5.0$      & $4.0$      & $3.0$      & $60$               \\
        Receiving Room   & $4.5$      & $4.0$      & $3.0$      & $54$               \\
        \hline
        \end{tabular}
\end{table*}
Next to this, the reverberation time $T$ in the rooms has to fulfil
\begin{equation}
    1 \leq T\leq 2\left(\frac{V}{50}\right)^{\frac{2}{3}}
\end{equation}
at $100\,\text{Hz}$, what will be realised by assuming a complex bulk modulus in Subsection \ref{Sec:VMFEM_AF}. \\
As already introduced in the definition of the STL in Equation (\ref{eqn:STL}) and Equation (\ref{eqn:EnerAveSP}), several measurement positions need to be defined to quantify the sound pressure level in the source and receiving room. Equivalently, source positions need to be defined in the source room. It should be mentioned that the DIN regulations also offer the possibility to excite and measure with moving sources resp. microphones - however, fixed source and microphone positions are straight-forward to realise in the FE model. Beginning with the source positions, the number of sound sources is determined by \cite{DIN10140-5}
\begin{equation}
    d = \frac{152}{V_\text{SR}^\frac{2}{3}},
\end{equation}
where $V_\text{SR}$ is the source room's volume. According to Table \ref{tab:DimBATF}, $d=10$ sources need to be placed; here, $12$ sound sources will be used to ensure a diffuse sound field. Thereby, a minimum distance of $0.7\,\text{m}$ and maximum distance of $1.4\,\text{m}$ between the sources have to be adhered. More precise, at least two sources must be no less than $1.4\,\text{m}$ distant from each other. Also, a minimum distance of $0.7\,\text{m}$ to the adjacent room limitations is defined. Furthermore, source positions must not be symmetric with respect to the mid-planes of the room, and source positions in the same plane must not be parallel but shifted with a distance of $0.1\,\text{m}$. According to the regulations, Table \ref{tab:SourcePos} summarises the source positions used in the subsequent contribution.
\begin{table*}[b!]
    \caption{Sound source positions in the source room.}\label{tab:SourcePos}
    \vspace{0.3cm}
    \centering
    \begin{tabular}{lcccccccccccc}
        \hline
        Source   &  $1$  &  $2$  &  $3$  &  $4$  &  $5$  & $6$   &  $7$  &  $8$  &  $9$  &  $10$ &  $11$ & $12$  \\
        \hline
        $x$ in m & $1.2$ & $0.8$ & $1.3$ & $1.5$ & $2.0$ & $2.2$ & $2.6$ & $2.8$ & $2.9$ & $3.2$ & $3.9$ & $4.0$ \\
        $y$ in m & $1.0$ & $3.0$ & $2.1$ & $1.8$ & $2.5$ & $1.2$ & $2.7$ & $1.6$ & $0.8$ & $2.4$ & $1.5$ & $3.1$ \\
        $z$ in m & $1.7$ & $1.2$ & $2.3$ & $0.9$ & $1.4$ & $2.0$ & $2.1$ & $1.8$ & $1.3$ & $1.1$ & $1.9$ & $0.8$ \\
        \hline
    \end{tabular}
\end{table*} \\
Next to the source positions defined by \cite{DIN10140-5}, the microphone positions follow from \cite{DIN10140-4}. Since more than one source is used, sound pressure level measurements must be taken at a minimum of five microphone positions; here, eight microphones will be utilised. When placing the microphones, there must be $0.7\,\text{m}$ distance to the adjacent microphones and room limitations; also $1\,\text{m}$ distance to the test specimen and sound sources. Moreover, two microphone positions must not lie in the same plane with respect to the room limitations. As a result to the previously listed regulations, the microphone positions in the source and receiving room are listed in Table \ref{tab:RecSRPos} and Table
\ref{tab:RecRRPos}.
\begin{table*}
    \caption{Microphone positions in the source room.}\label{tab:RecSRPos}
    \vspace{0.3cm}
    \centering
    \begin{tabular}{lcccccccc}
        \hline
        Microphone &  $1$  &  $2$  &  $3$  &  $4$  &  $5$  & $6$   &  $7$  &  $8$  \\
        \hline
        $x$ in m   & $0.7$ & $1.2$ & $2.0$ & $1.9$ & $2.6$ & $3.4$ & $4.0$ & $3.5$ \\
        $y$ in m   & $1.6$ & $2.5$ & $1.0$ & $3.0$ & $2.2$ & $0.8$ & $1.7$ & $3.3$ \\
        $z$ in m   & $1.4$ & $1.6$ & $1.0$ & $2.0$ & $0.9$ & $2.3$ & $0.7$ & $1.6$ \\
        \hline
    \end{tabular}
\end{table*}
\begin{table*}
    \caption{Microphone positions in the receiving room, where the offset $l_{x_\text{SR}}+l_{x_\text{G}}$ needs to be considered.}\label{tab:RecRRPos}
    \vspace{0.3cm}
    \centering
    \begin{tabular}{lcccccccc}
        \hline
        Microphone &  $1$  &  $2$  &  $3$  &  $4$  &  $5$  & $6$   &  $7$  &  $8$  \\
        \hline
        $x$ in m   & $1.2$ & $1.4$ & $1.6$ & $1.9$ & $2.6$ & $2.7$ & $3.6$ & $3.3$ \\
        $y$ in m   & $1.3$ & $3.0$ & $2.3$ & $1.7$ & $2.6$ & $1.4$ & $1.0$ & $3.0$ \\
        $z$ in m   & $0.8$ & $1.6$ & $1.1$ & $2.1$ & $1.4$ & $1.7$ & $2.2$ & $1.0$  \\
        \hline
    \end{tabular}
\end{table*}
In Table \ref{tab:RecRRPos}, an offset $l_{x_\text{SR}}+l_{x_\text{G}}$ needs to be considered, e.g., the microphone position $1$ in the receiving room is in fact $x=6.2\,\text{m}+l_{x_\text{G}}$, while $l_{x_\text{G}}$ will be specified in Section \ref{Sec:BATF_FEM}. Note that the walls are not included into the offset as they will be modelled as 2D continua in Subsection \ref{Sec:VMFEM_STR}. Lastly, it is to mention that not all source and microphone positions fulfil the regulations regarding minimum distances to each other exactly; however, the majority does.\\
Having set up the building acoustics test facility according to DIN EN ISO 10140 that will be utilised in this contribution, the following section introduces the physical modelling of airborne and structure-borne sound in the different domains, as well as the resulting FE formulations.

\section{Vibroacoustic Model and Finite Element Formulations}\label{Sec:VMFEM}
Summarising the vibroacoustic chain of the building acoustics test facility with a double-leaf wall test specimen already introduced in Section \ref{Sec:BATF}, airborne sound radiates from the sound sources into the source room, which then excites the leaf facing the source room to structure-borne sound. The leaf excites the gap-filling air to airborne sound or the gap-filling porous insulation material to airborne and structure-borne sound that consequently loads the leaf facing the receiving room. Finally, airborne sound radiates into the receiving room. In the following, Subsection \ref{Sec:VMFEM_AF} introduces the airborne sound propagation in the source and receiving room as well as the air-filled gap, while Subsection \ref{Sec:VMFEM_EQF} describes the sound propagation within the insulation using an equivalent fluid approach that is briefly introduced there. Then, Subsection \ref{Sec:VMFEM_STR} addresses structure-borne sound propagation in plane-shell-like structures and, lastly, Subsection \ref{Sec:VMFEM_FSI} the coupling of (equivalent) fluid and structure.

\subsection{Acoustic Fluid}\label{Sec:VMFEM_AF}
Sound propagation in fluids is described by longitudinal waves. Here, a bounded fluid domain is considered, since the sound propagation takes place in two rooms - and the gap, if no insulation is applied. The partial differential equation (PDE) in the time-domain describing the sound pressure $p(\mathbf{x},t)$ in a perfect fluid domain $\Omega_a$ due to a monopole source with volume velocity $Q_s$ per unit volume at position $\mathbf{x}_s\in\Omega_a$ can be derived, for instance, using a variational formulation arising from Hamilton’s principle. Then, assuming a stationary harmonic state, the Helmholtz equation follows as \cite{FAH2003}
\begin{equation}\label{eqn:Helm}
    \nabla^2\hat{p}(\mathbf{x})+k^2\hat{p}(\mathbf{x})=-i\omega\rho_0 Q_s\delta(\mathbf{x}-\mathbf{x}_s).
\end{equation}
Therein, $\hat{p}(\mathbf{x})$ is the sound pressure in the perfect fluid domain $\Omega_a$, $k=\omega/c$ the wave number, $\omega$ the angular frequency, $c$ the speed of sound, $\rho_0$ the fluid density, $\delta(\mathbf{x}-\mathbf{x}_s)$ the Dirac delta distribution, and $i$ the imaginary unit. Together with three different boundary conditions on the fluid domain's boundary $\partial\Omega_a$, Equation (\ref{eqn:Helm}) defines the acoustic boundary value problem \cite{ATA2015}:
\begin{itemize}
    \item Dirichlet boundary condition on $\partial\Omega_{a,D}$: prescribed sound pressure 
        \begin{equation}
            \hat{p}(\mathbf{x})=\hat{\overline{p}}(\mathbf{x}),\,\forall\,\mathbf{x}\in\partial\Omega_{a,D};
        \end{equation}
    \item Neumann boundary condition on $\partial\Omega_{a,N}$: prescribed normal sound flux 
        \begin{equation}
            \frac{\partial\hat{p}(\mathbf{x})}{\partial \mathbf{n}}=\rho_0\omega \hat{v}(\mathbf{x})\mathbf{n},\,\forall\,\mathbf{x}\in\partial\Omega_{a,N},
        \end{equation}
    where $\hat{v}$ is the particle velocity and $\mathbf{n}$ is the surface's normal vector;
    \item mixed boundary condition on $\partial\Omega_{a,m}$: prescribed specific normalised acoustic admittance 
        \begin{equation}
            \frac{\partial\hat{p}(\mathbf{x})}{\partial \mathbf{n}} = -ik\frac{\rho_0c_0}{Z_n}\hat{p}(\mathbf{x})\,\forall\,\mathbf{x}\in\partial\Omega_{a,m},
        \end{equation}
    where $Z_n$ is the specific acoustic impedance on $\partial\Omega_{a,m}$.
\end{itemize}
In the following, damping in the fluid is modelled by assuming a complex bulk modulus $\hat{K}$ resulting in a complex speed of sound $\hat{c}=c(1+i\eta_a)$, where $\eta_a$ is the fluid's loss factor \cite{ATA2015}\cite{KLI2008}:
\begin{equation}\label{eqn:LossFactorsAir}
    \eta_a=\frac{2.2}{fT}=\frac{2.2c}{6\ln(10)}\frac{m}{f},
\end{equation}
with $T$ as the reverberation time, $m$ the propagation loss in air, and $f$ the frequency. The propagation loss $m$ is calculated as described in \cite{BAS1990}, \cite{BAS1995} and \cite{ISO9613}.\\
The PDE in Equation (\ref{eqn:Helm}) defines the strong formulation; however, the differentiability requirements need to be reduced, which is why a weak formulation is derived. This can be done using, for instance, again a variational formulation or the Galerkin method \cite{ATA2015}. Then, the domain is discretised using shape functions, followed by the approximation of the test functions and physical quantities \cite{ATA2015}. Here, the same shape functions are used for the discretisation of the geometry as well as the approximation of the test functions and physical quantities - i.e., isoparametric elements. \\
Finally, the element matrices are assembled to the global ones and the Helmholtz equation's FE formulation in matrix form reads 
\begin{equation}\label{eqn:HelmFEM}
    \left(\frac{1}{\rho_0\omega^2}\hat{\mathbf{K}}_a-\frac{1}{\rho_0\hat{c}^2}\hat{\mathbf{M}}_a\right)\hat{\mathbf{p}}_a=i\frac{Q_s}{\omega}\mathbf{w},
\end{equation}
where $\hat{\mathbf{K}}_a$, $\hat{\mathbf{M}}_a\in\mathbb{C}^{n_a\times n_a}$ are the stiffness and mass matrix, $\hat{\mathbf{p}}_a\in\mathbb{C}^{n_a}$ the vector of nodal sound pressure, and $\mathbf{w}\in\mathbb{R}^{n_a}$ a vector, which entry $w_i=1$ at the $i$-th node where the monopole source is applied and elsewhere zero. The spatial dependencies are neglected in Equation (\ref{eqn:HelmFEM}) due to readability reasons. Moreover, the FE formulation is scaled with $1/\omega^2$ to preserve the symmetry of the linear system of equation when coupling fluid and structure (see Subsection \ref{Sec:VMFEM_FSI}).

\subsection{Equivalent Fluid}\label{Sec:VMFEM_EQF}
In general, porous materials are suitable materials for sound insulations, as they transfer sound-wave-induced motions into heat \cite{BOL1997}. A porous medium consists of an elastic frame filled with a fluid, here air. Due to the interaction between the fluid and elastic frame, the sound propagation is a coupled problem that can be described using the Biot theory \cite{ALL2009}. However, neglecting the elasticity of the frame and assuming an rigid frame, the sound propagation through porous media is described by an equivalent fluid approach. To use the equivalent fluid approach, the long-wavelength condition must be fulfilled; i.e. the characteristic dimensions of the pores are much smaller compared to the wavelength, and the fluid is incompressible at microscopic scale \cite{ALL2009}. It should be evident that disregarding the elasticity of the frame yields a model error especially when the insulation is placed between two oscillating surfaces - this is described in \cite{ALL2009} and documented for a double-leaf wall in \cite{CLA2008}. Nevertheless, the simplicity of implementing the equivalent fluid approach is appealing since the sound propagation in the insulation is still described by the Helmholtz equation (Equation (\ref{eqn:Helm})). \\
According to the Johnson-Champoux-Allard (JCA) model \cite{ALL2009}, the bulk modulus $K$ and density $\rho_0$ are simply replaced by a complex effective bulk modulus $\hat{K}_e$ 
\begin{equation}
    \hat{K}_e = \gamma P_0\left[ \gamma-\frac{\gamma-1}{1+\frac{8\mu}{i\Lambda'^2\text{Pr}\rho_0\omega}\sqrt{1+\frac{i\Lambda'^2\text{Pr}\rho_0\omega}{16\mu}}} \right]^{-1},
\end{equation}
and complex effective density $\hat{\rho}_e$
\begin{equation}
    \hat{\rho}_e= \alpha_\infty\rho_0 \left[ 1 + \frac{\sigma\phi}{i\alpha_\infty\rho_0\omega}\sqrt{1+\frac{i4\alpha_\infty^2\mu\rho_0\omega}{\sigma^2\phi^2\Lambda^2}}\right].
\end{equation}
Here, $\alpha_\infty$ is the tortuosity, $\gamma$ the heat capacity ratio, $\mu$ the viscosity, $\Lambda$ the viscous characteristic length, $\Lambda'$ the thermal characteristic length, $\sigma$ the flow resistivity, $\phi$ the porosity, $P_0$ the ambient pressure, and $\text{Pr}$ the Prandtl number. Using the relation $K=\rho_0c^2$, the complex speed of sound in the porous media is \cite{FAH2003}
\begin{equation}
    \hat{c}_e=\sqrt{\frac{\hat{K}_e}{\hat{\rho}_e}}.
\end{equation}
At least, to consider the mass inertia of the frame at low frequencies, the assumption of an rigid frame can be replaced by the assumption of a limp frame \cite{ALL2009}. The equations behind the limp approach can be derived from the Biot theory neglecting the stress tensor of the solid phase (frame) yielding, again, the Helmholtz equation, but now incorporating the effective density for a limp material $\hat{\rho}_{e,\text{limp}}$. According to \cite{PAN2007}, the effective density is
\begin{equation}
    \hat{\rho}_{e,\text{limp}}\approx\frac{\hat{\rho}_eM-\rho_0^2}{M+\hat{\rho}_e-2\rho_0},    
\end{equation}
and $M=\rho_1+\phi\rho_0$ is the total apparent mass of the equivalent fluid limp medium with the in vacuo bulk density $\rho_1$ of the medium. It should be mentioned that the limp approach applies to low-density porous media (e.g. aeronautic grade fiberglass); nonetheless, for heavy frames, the effective density $\hat{\rho}_{e,\text{limp}}$ converges against the effective density $\hat{\rho}_e$ \cite{ALL2009}\cite{PAN2007}. \\
Consequently, the FE formulation of sound propagation in the equivalent fluid assuming a limp frame, that is not loaded with a source term, reads
\begin{equation}\label{eqn:EqFlFEM}
    \left(\frac{1}{\hat{\rho}_{e,\text{limp}}\omega^2}\hat{\mathbf{K}}_e-\frac{1}{\hat{\rho}_{e,\text{limp}}\hat{c}_e^2}\hat{\mathbf{M}}_e\right)\hat{\mathbf{p}}_e=\mathbf{0},
\end{equation}
where $\mathbf{\hat{K}}_e$, $\mathbf{\hat{M}}_e\in\mathbb{C}^{n_e\times n_e}$ are the stiffness and mass matrix, and $\hat{\mathbf{p}}_e\in\mathbb{C}^{n_e}$ the vector of nodal sound pressure. Hence, the use of the equivalent fluid approach will not introduce new degrees of freedom as well as new formulations of the coupling conditions into the FE formulation of the coupled vibroacoustic problem (see Subsection \ref{Sec:VMFEM_FSI}).

\subsection{Structure}\label{Sec:VMFEM_STR}
In contrast to fluids, solid structures are capable of resisting shear stresses which yields different wave types describing the solids' oscillation next to the already known longitudinal waves. In acoustics, the bending wave is of utmost importance in the context of sound radiation of oscillating structures as it describes the structures' out-of-plane movement \cite{MOE2010}. However, in building acoustics, also in-plane oscillations are important to be considered when modelling flanking transmission where two walls are connected at an corner. This contribution neglects flanking transmission; however, the in-plane movement will be included into the finite element formulation since it represents the general building acoustics use case. \\
In the following, the three-dimensional solid walls will be abstracted as a two-dimensional structure (length, width $\gg$ thickness) which offers the possibility to describe the wall's bending (out-of plane) oscillations based on the Mindlin plate theory and the in-plane oscillations based on a membrane theory - overall, this yields a plane shell formulation. The Mindlin theory is used to ensure that also thick walls ($\text{thickness}/\text{min}(\text{length,width}) > 0.2$) can be modelled with sufficient accuracy \cite{ALT2023}. Hence, the main assumptions are finite shear stiffness as well as plane stress and no deformations of the cross-sections \cite{ALT2023}. Then, the kinematic variables describing the bending and in-plane waves are $\mathbf{u}=[u_{1M},u_{2M},u_3,\varphi_1,\varphi_2]^T$, where $u_{1M}$, $u_{2M}$ are the in-plane displacements of the mid-plane, $u_3$ the out-of-plane displacement, and $\varphi_1$, $\varphi_2$ the rotations around the in-plane axes. \\
Similar to the sound pressure field $p(\mathbf{x})$ in Subsection \ref{Sec:VMFEM_AF}, the PDEs describing the kinematic variables $\mathbf{u}$ of a plane shell occupying the domain $\Omega_s$ can be derived using, for instance, Hamilton's principle. Assuming, again, a stationary harmonic state, the PDEs are \cite{CLA2008}\cite{MOE2010}
\begin{flalign}\label{eqn:PlShell_PDE_01}
    & \frac{Eh}{1-\nu^2}\hat{u}_{1M,11}+Gh\left(\hat{u}_{1M,22}+\frac{1+\nu}{1-\nu}\hat{u}_{2M,12}\right) + \omega^2\rho h\hat{u}_{1M} = -\hat{p}_1,
\end{flalign}  
\begin{flalign}\label{eqn:PlShell_PDE_02}
    & \frac{Eh}{1-\nu^2}\hat{u}_{2M,22}+Gh\left(\hat{u}_{2M,11}+\frac{1+\nu}{1-\nu}\hat{u}_{1M,12} \right) + \omega^2\rho h\hat{u}_{2M}=-\hat{p}_2,
\end{flalign}
\begin{flalign}\label{eqn:PlShell_PDE_03}
    & Ghk_s\left(\Delta \hat{u}_3+\hat{\varphi}_{1,1}+\hat{\varphi}_{2,2} \right) +\omega^2\rho h\hat{u}_3 = -\hat{p}_3, 
\end{flalign}  
\begin{flalign}\label{eqn:PlShell_PDE_04}
    & GI\hat{\varphi}_{1,22}+B\hat{\varphi}_{1,11}+\frac{1+\nu}{2}B\hat{\varphi}_{2,12}-Ghk_s\left(\hat{u}_{3,1}+\hat{\varphi}_1 \right) + \omega^2\rho I\hat{\varphi}_1 = 0, 
\end{flalign}  
\begin{flalign}\label{eqn:PlShell_PDE_05}
    & GI\hat{\varphi}_{2,11}+B\hat{\varphi}_{2,22}+\frac{1+\nu}{2}B\hat{\varphi}_{1,12}-Ghk_s\left(\hat{u}_{3,2}+\hat{\varphi}_2 \right) + \omega^2\rho I\hat{\varphi}_2 = 0, 
\end{flalign}
where $B=EI/(1-\nu^2)$ is the bending stiffness, $E$ Young's modulus, $G=E/(2(1+\nu))$ the shear stiffness, $h$ the plate thickness, $I=h^3/12$ the geometrical moment of interia of a plate, $k_s=5/6$ the shear correction factor, $p_i$ the loads in coordinate directions $i=1,2,3$, $\nu$ Poisson's ratio, and $\rho$ the plate density. The index after the comma indicates the partial derivative in the $i$-th coordinate direction, $i=1,2,3$. As can be seen in the equations above, the membrane's in-plane displacements $u_{1M}$, $u_{2M}$ are not coupled to the plate's out-of-plane displacement $u_3$ and rotations $\varphi_1$, $\varphi_2$. This assumption is valid for plane shells. \cite{ZIE2005}. \\
Now, defining boundary conditions for the boundary $\partial\Omega_s$ of the plane shell domain $\Omega_s$, the PDEs in Equation (\ref{eqn:PlShell_PDE_01}) to Equation (\ref{eqn:PlShell_PDE_05}) in combination with the respective boundary conditions describe a structural boundary value problem \cite{ATA2015}\cite{ZIE2005}:
\begin{itemize}
    \item Dirichlet boundary conditions on $\partial\Omega_{s,D}$: prescribed displacements and rotations
        \begin{equation}
            \hat{\mathbf{u}}(\mathbf{x}) = \hat{\overline{\mathbf{u}}}(\mathbf{x}),\,\forall\mathbf{x}\in\partial\Omega_{s,D};
        \end{equation}
    \item Neumann boundary condition on $\partial\Omega_{s,N}$: prescribed stress resultants, i.e. shear forces, axial forces and moments;
    \item mixed boundary condition on $\partial\Omega_{s,m}$: combination of Dirichlet and Neumann boundary condition.
\end{itemize}
Structural damping is assumed similarly to the fluid by defining a complex Young's modulus $\hat{E}=E(1+i\eta_s)$, where $\eta_s$ is the structure's loss factor \cite{ATA2015}. Typically, $\eta_s\in[0.02,0.05]$ is valid for common building materials \cite{KLI2008}. \\
Subsequently, the assembled FE formulations of the plane shell can be derived equivalently to the procedure described in Subsection \ref{Sec:VMFEM_AF}. Nonetheless, special attention must be paid to the assembly of the global FE matrices and vectors. The element matrices are derived for an plane shell element living in a local coordinate system in the physical space and, therefore, need to be transformed into the global coordinate system of the physical space. A simple transformation approach includes the construction of a projection matrix, whose entries are the cosines between the global and local coordinate axes. Yet the transformation affects the three main axes of the coordinate system, but the vector of the kinematic variables $\mathbf{u}$ only contains five of six degrees of freedom. Hence, the missing rotational degree of freedom $\varphi_3$ needs to be added, which is done by assuming an artificial stiffness \cite{CLA2008}\cite{ZIE2005}. There exist also other approaches to include the missing degree of freedom that are referenced in \cite{ZIE2005}. \\
It is also known that plate and membrane elements suffer from locking effects introduced by parasitic stresses. For the here discussed use cases, the plane shell element's plate part could suffer from locking due to the leafs' thickness \cite{CLA2008}; as a solution, the discrete shear gap (DSG) method as described in \cite{BLE2000} is applied to the plate part, which aims to reduce shear locking effects occurring in thin Mindlin plates. This approach effects the shape functions in the FE formulation describing the shear effects.\\
Consequently, the FE formulation for the plane shell element in local coordinates is \cite{CLA2008}
\begin{flalign}
    &\left(
        \begin{bmatrix}
            \hat{\mathbf{K}}^P& \mathbf{0}    & \mathbf{0} \\
            \mathbf{0}  & \hat{\mathbf{K}}^M  & \mathbf{0} \\
            \mathbf{0}  & \mathbf{0}    & \hat{\mathbf{K}}_{\varphi_3}
        \end{bmatrix}
        - \omega^2
        \begin{bmatrix}
            \hat{\mathbf{M}}^P& \mathbf{0}    & \mathbf{0} \\
            \mathbf{0}  & \hat{\mathbf{M}}^M  & \mathbf{0} \\
            \mathbf{0}  & \mathbf{0}    & \hat{\mathbf{M}}_{\varphi_3}^P
        \end{bmatrix}
    \right)
    \begin{bmatrix}
        \hat{\mathbf{u}}^P \\
        \hat{\mathbf{u}}^M \\
        \hat{\mathbf{\varphi}}_3
    \end{bmatrix} 
    =
    \begin{bmatrix}
        \hat{\mathbf{f}}^P \\
        \hat{\mathbf{f}}^M \\
        \mathbf{0}
    \end{bmatrix},
\end{flalign}
where the superscripts $P$ (plate), $M$ (membrane), and subscript $\varphi_3$ label the respective stiffness matrices $\hat{\mathbf{K}}$, mass matrices $\hat{\mathbf{M}}$, vectors of nodal displacements $\hat{\mathbf{u}}$, and load vectors $\hat{\mathbf{f}}$ \cite{CLA2008}. Finally, applying the transformation to the global coordinate system and assembling the element matrices yields the global FE formulation 
\begin{equation}
    \left(\hat{\mathbf{K}}_s -\omega\hat{\mathbf{M}}_s\right)\hat{\mathbf{u}}_s = \hat{\mathbf{f}}_s,
\end{equation}
with $\hat{\mathbf{K}}_s$, $\hat{\mathbf{M}}_s\in\mathbb{C}^{n_s\times n_s}$, $\hat{\mathbf{u}}_s\in\mathbb{C}^{n_s}$, and $\hat{\mathbf{f}}_s\in\mathbb{C}^{n_s}$.

\subsection{Fluid-Structure Coupling}\label{Sec:VMFEM_FSI}
Generally, two different approaches exist to describe the coupling between fluid and structure domains: weak and strong coupling. In the case of weak coupling, the sound pressure load onto the structure is negligible and the subdomains can be solved subsequently. Here, strong coupling is considered, where the different domains are solved simultaneously, as the sound pressure's load onto the structure is considered \cite{KAL2015}. \\
Since the insulation is modelled as an equivalent acoustic fluid, only fluid-structure coupling is required in the following. Then, the continuity condition at the fluid-structure-interface demands that the normal component of the structure's surface velocity equals the normal component of the fluid's particle velocity \cite{ATA2015}\cite{KAL2015}. Utilising the linear momentum, the relation between the sound pressure $\hat{p}$ and displacement $\hat{\mathbf{u}}$ is \cite{ATA2015}
\begin{equation}
    \frac{\partial\hat{p}}{\partial\mathbf{n}} = \rho_0\omega^2\hat{\mathbf{u}}\cdot\mathbf{n}.
\end{equation}
Furthermore, the stress-continuity at the fluid-structure-interface requires \cite{ATA2015}
\begin{equation}
    \boldsymbol{\sigma}\cdot\mathbf{n} = \hat{p}\mathbf{n}.
\end{equation}
Equivalently, when using Hamilton's principle, the coupling conditions describe the coupling as an external work of the structure on the fluid and vice versa. \\
Subsequently, the coupling conditions introduce coupling matrices $\mathbf{\hat{C}}$ into the FE formulation, which then reads for the building acoustics test facility consisting of three fluid domains and two structural domains
\begin{flalign}
    \left(\hat{\mathbf{K}}-\omega^2\hat{\mathbf{M}}\right)\hat{\mathbf{x}} = \hat{\mathbf{f}},
\end{flalign}
where 
\begin{flalign}
    \hat{\mathbf{K}}= 
        \begin{bmatrix}
            \frac{1}{\rho_0\omega^2}\hat{\mathbf{K}}_\text{G}    & \mathbf{0}                                    & \mathbf{0}                                    & \mathbf{0}                & \mathbf{0} \\
            \mathbf{0}                                      & \frac{1}{\rho_0\omega^2}\hat{\mathbf{K}}_\text{RR} & \mathbf{0}                                    & \mathbf{0}                & \mathbf{0} \\
            \mathbf{0}                                      & \mathbf{0}                                    & \frac{1}{\rho_0\omega^2}\hat{\mathbf{K}}_\text{SR} & \mathbf{0}                & \mathbf{0}\\
            -\mathbf{\hat{C}}                               & -\hat{\mathbf{C}}                             & \mathbf{0}                                    & \hat{\mathbf{K}}_{\text{W}_\text{RR}} & \mathbf{0} \\
            -\hat{\mathbf{C}}                               & \mathbf{0}                                    & -\hat{\mathbf{C}}                             & \mathbf{0}                & \hat{\mathbf{K}}_{\text{W}_\text{SR}} 
        \end{bmatrix} \notag
\end{flalign}
is the global stiffness matrix ($\mathbf{\hat{K}}\in\mathbb{C}^{n\times n}$),
\begin{flalign}
        \hat{\mathbf{M}}= 
            \begin{bmatrix}
            \frac{1}{\rho_0\hat{c}^2\omega^2}\hat{\mathbf{M}}_\text{G}   & \mathbf{0}                                            & \mathbf{0}                                            & \frac{1}{\omega^2}\hat{\mathbf{C}}^T  & \frac{1}{\omega^2}\hat{\mathbf{C}}^T \\
            \mathbf{0}                                              & \frac{1}{\rho_0\hat{c}^2\omega^2}\hat{\mathbf{M}}_\text{RR}& \mathbf{0}                                            & \frac{1}{\omega^2}\hat{\mathbf{C}}^T  & \mathbf{0} \\
            \mathbf{0}                                              & \mathbf{0}                                            & \frac{1}{\rho_0\hat{c}^2\omega^2}\hat{\mathbf{M}}_\text{SR}& \mathbf{0}                            & \frac{1}{\omega^2}\hat{\mathbf{C}}^T\\
            \mathbf{0}                                              & \mathbf{0}                                            & \mathbf{0}                                            & \hat{\mathbf{M}}_{\text{W}_\text{RR}}             & \mathbf{0} \\
            \mathbf{0}                                              & \mathbf{0}                                            & \mathbf{0}                                            & \mathbf{0}                            & \hat{\mathbf{M}}_{\text{W}_\text{SR}} 
            \end{bmatrix} \notag
\end{flalign}
the global mass matrix ($\mathbf{\hat{M}}\in\mathbb{C}^{n\times n}$),
\begin{flalign}
    \hat{\mathbf{x}}^T =
        \begin{bmatrix}
            \hat{\mathbf{p}}_\text{G} & \hat{\mathbf{p}}_\text{RR} & \hat{\mathbf{p}}_\text{SR} & \hat{\mathbf{u}}_{\text{W}_\text{RR}} & \hat{\mathbf{u}}_{\text{W}_\text{SR}} 
        \end{bmatrix}^T \notag
\end{flalign}
the global vector of nodal degrees of freedom ($\hat{\mathbf{x}}\in\mathbb{C}^n$), and
\begin{flalign}
    \hat{\mathbf{f}}^T =
        \begin{bmatrix}
            \mathbf{0} & \mathbf{0} & i\frac{Q_s}{\omega}\mathbf{w} &\mathbf{0} &\mathbf{0} \\
        \end{bmatrix}^T \notag
\end{flalign}
the global load vector ($\hat{\mathbf{f}}\in\mathbb{C}^n$). The subscripts G, SR, RR refer to the three fluid domains gap, source room, and receiving room; while W refers to the respective walls. It should be mentioned that there only exists an external load due to the monopole sources in the source room. Moreover, the stiffness and mass matrices for the fluid in the gap as defined above consider only atmospherically damped air. If an equivalent fluid shall be considered, $\rho_0$ changes to $\hat{\rho}_{e,\text{limp}}$ and $\hat{c}$ to $\hat{c}_e$. \\
\\
Lastly, a comment regarding conforming and non‑conforming meshes is given at this point, since it affects the construction of the coupling matrices. Generally, the length of the finite elements is determined by the wavelengths of the sound wave in the respective domain at the maximum frequency of interest. The famous rule of thumb states that at least six resp. four elements per wavelength are required for linear resp. quadratic elements to yield a sufficient resolution of acoustic wave phenomena using FEM \cite{ATA2015}. Transferred to nodes per wavelength, seven resp. nine nodes are required for linear resp. quadratic elements. Here, it is to be expected that the wavelength and, thus, the required minimal element length differ for the different domains due to the different wave types and materials in the different domains. \\ 
Now, using a geometrically conforming mesh, where the same element length is used in all domains, which needs to be chosen with respect to the overall smallest required element length of the different domains, the coupling conditions will be fulfilled node-wise at the domain interfaces since there exists a one-to-one correspondence of the respective interface nodes. However, this may yield a time and resource consuming meshing and solving procedure depending on the minimal required element length. To overcome this bottleneck, a domain-specific discretisation approach, cf. \cite{BLE2024}, based on non-conforming meshes can be used, since then the domains can be meshed with regard to the respective minimum required element length. Nonetheless, there no longer exists a one-to-one correspondence between the nodes of the different domains at an interface; hence, special attention has to be paid to the construction of the coupling matrices. Here, a Nitsche-type coupling procedure is used as described in \cite{KAL2015}.\\
\\
Having introduced the FE formulations of the different test facility domains - namely the source and receiving room, the air-filled or insulation-filled gap, and the double-leaf wall - as well as the fluid-structure coupling, the subsequent section presents the key facts of the building acoustics test facility FE model.

\section{Finite Element Model of the Building Acoustic Test Facility}\label{Sec:BATF_FEM}
\begin{figure}[!b]
    \centering
    \def\svgwidth{\linewidth}
    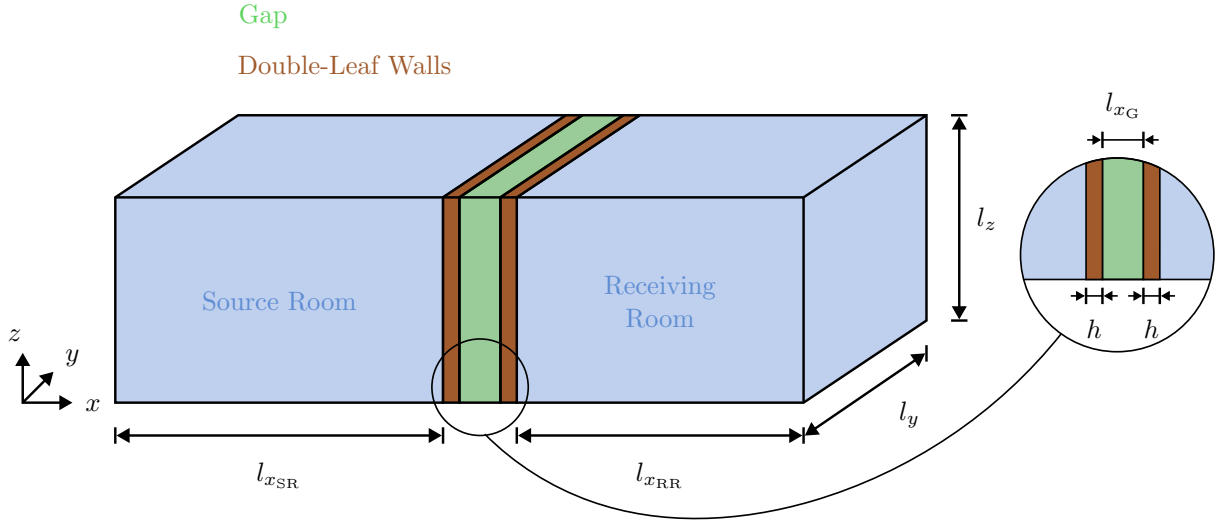
    \caption{Schematic sketch of the building acoustics test facility FE Model consisting of a source and receiving room as well as a double-leaf wall as a test specimen. The measures are in compliance with Table \ref{tab:DimBATF}, while $l_{x_\text{G}}=0.05\,\text{m}$ and $h=0.025\,\text{m}$.}
    \label{fig:BATF_FEM}
\end{figure}

\begin{figure}[b!]
    \centering
        \begin{tikzpicture}
            \begin{semilogyaxis}[
                xlabel={Frequency in Hz},
                ylabel={Loss Factor},
                xmin=100, xmax=3150,
                width=12cm,
                height=4.5cm,
                scaled y ticks=true,
                legend style={
                at={(0.5,1.2)},
                anchor=center,
                draw=black,
                legend cell align=left,
                fill=none,
                legend columns=3,
                font=\small
                }
                ]
                \addplot [mark=none, color=spec_blue, thick] table [
                    x index=0, 
                    y index=1
                    ] {eta_WTF_Trev.txt};
                \addlegendentry{$\eta_{a,T}$}
                \addplot [mark=none, color=spec_green, thick] table [
                    x index= 0, 
                    y index= 1 
                    ] {eta_WTF_Atmospheric.txt};
                \addlegendentry{$\eta_{a,m}$}
                \addplot [mark=none, color=spec_brown, thick] table [
                    x index=0, 
                    y index=1
                    ] {eta_WTF_Walls.txt};
                \addlegendentry{$\eta_s$}
            \end{semilogyaxis}
        \end{tikzpicture}
    \caption{Total loss factor $\eta_{a,T}$ applied to the fluid in the source and receiving room considering a reverberation time $T=1.5\,\text{s}$; total loss factor $\eta_{a,m}$ applied to the fluid in the gap, if no insulation is applied, considering the propagation loss $m$; as well as the total loss factor $\eta_s$ of the structure displayed over the building-acoustics-relevant frequency range $100\,\text{Hz}$ to $3150\,\text{Hz}$.}
    \label{fig:eta}
\end{figure}
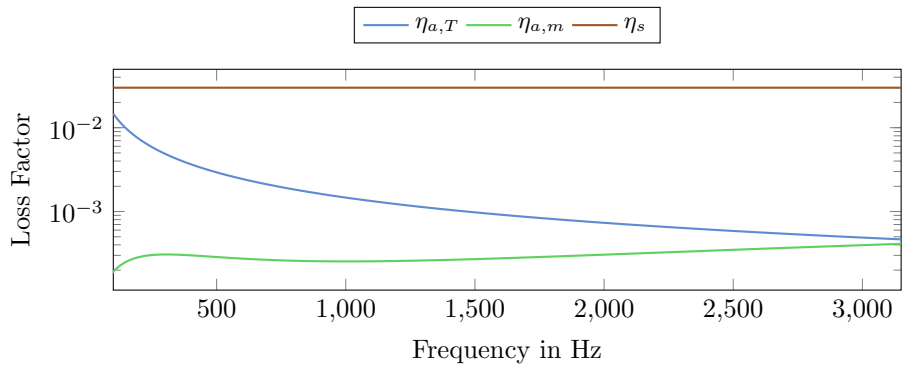

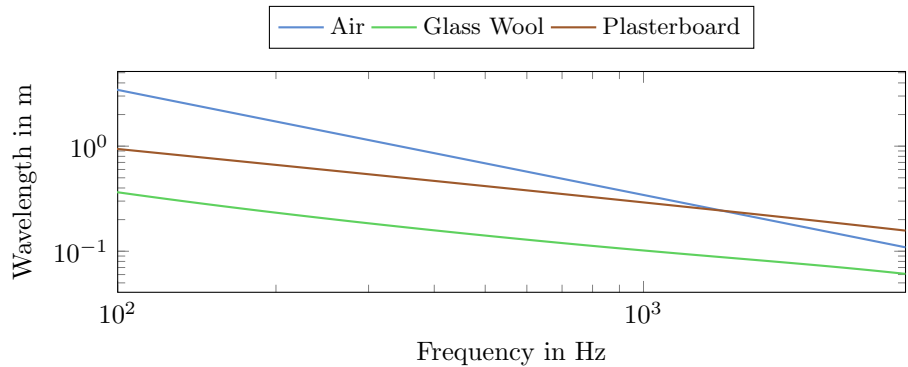
\begin{figure}[b!]
    \centering
        \begin{tikzpicture}
            \begin{loglogaxis}[
                 xtick={
                    100,
                    1000},
                xlabel={Frequency in Hz},
                ylabel={Wavelength in m},
                xmin=100, xmax=3150,
                width=12cm,
                height=4.5cm,
                scaled y ticks=true,
                legend style={
                at={(0.5,1.2)},
                anchor=center,
                draw=black,
                legend cell align=left,
                fill=none,
                legend columns=3,
                font=\small
                }
                ]
                \addplot [mark=none, color=spec_blue, thick] table [
                    x index= 0, 
                    y index= 1 
                    ] {wavenlength_3150Hz_fluid.txt};
                \addlegendentry{Air}
                \addplot [mark=none, color=spec_green, thick] table [
                    x index=0, 
                    y index=1
                    ] {wavenlength_3150Hz_equivalentFluid.txt};
                \addlegendentry{Glass Wool}
                \addplot [mark=none, color=spec_brown, thick] table [
                    x index=0, 
                    y index=1
                    ] {wavenlength_3150Hz_structure.txt};
                \addlegendentry{Plasterboard}
            \end{loglogaxis}
        \end{tikzpicture}
    \caption{Wavelength of the longitudinal resp. bending waves in air (fluid), glass wool (equivalent fluid) and plasterboard (structure) displayed over the building-acoustics-relevant frequency range $100\,\text{Hz}$ to $3150\,\text{Hz}$.}
    \label{fig:waveLengths}
\end{figure}

Figure \ref{fig:BATF_FEM} illustrates the building acoustics test facility FE model, which neglects, compared to Figure \ref{fig:BATF}, the flanking walls since flanking transmission is not part of the following investigations. 
As a consequence, the plane shells' in-plane degrees of freedom will not contribute to the FEM solution; however, the formulation of a plane shell element is useful for future investigations considering flanking transmission, which is why the plane shell element will be used throughout the following investigations. Moreover, compared to \cite{ARJ2014}, connections between the leafs of the double-leaf wall are not modelled in this FE model; thus, the leafs are only connected via the air spring defined by the properties of air or the equivalent fluid. The distance between the double-leaf walls is $l_{x_\text{G}}=0.05\,\text{m}$, while a leaf's thickness measures $h=0.025\,\text{m}$. The remaining measures of the test facility correspond to those in Table \ref{tab:DimBATF}. Furthermore, all boundary degrees of freedom of the walls are set to zero - i.e., the walls are fully clamped (Dirichlet boundary condition) - while the boundaries of the fluid domains are perfectly reflecting (Neumann boundary condition) what is already implicitly fulfilled by the weak integral formulation of the Helmholtz equation in Subsection \ref{Sec:VMFEM_AF}. \\
Subsequently, the following materials will be used:
\begin{itemize}
    \item Air ($T=293.15\,\text{K}$, $P_0=101325\,\text{Pa}$,  $c =343\,\text{m/s}$, $\rho_0=1.2041\,\text{kg/m}^3$, $\gamma=1.4$, $\mu=18.1e{-6}\,\text{Ns/m}^2$, $Pr=0.7039$, relative humidity $h_r=40\,\text{\%}$) \cite{BAE2016}\cite{DIE2024}
    \item Glass wool ($\alpha_\infty=1.06$, $\Lambda=56e{-6}\,\text{m}$, $\Lambda'=110e{-6}\,\text{m}$, $\sigma=40000\,\text{Ns/m}^4$, $\phi=0.94$, $\rho_1=130\,\text{kg/m}^3$) \cite{ATA1998}
    \item Plasterboard ($E=3e9$\,\text{Pa}, $\nu=0.15$, $\rho=800\,\text{kg/m}^3$) \cite{CLA2008}
\end{itemize}
Accordingly, air is used within the source and receiving room as well as in the gap resp. the insulation's fluid phase. In the case of the source and receiving room, the loss factor $\eta_a$ is defined by the reverberation time $T=1.5\,\text{s}$ in compliance with DIN EN ISO 10140-5 - in the following $\eta_{a,T}$. If no insulation is placed in the gap, the loss factor follows from the propagation loss $m$ - i.e., atmospheric damping is assumed - in the following $\eta_{a,m}$. If an insulation is placed within the gap, then it is glass wool. Figure \ref{fig:eta} show both loss factors in the building-acoustics-relevant frequency range $100\,\text{Hz}$ to $3150\,\text{Hz}$. As can be seen, both are frequency-dependent. 
Lastly, the double-leaf wall consists of plasterboard. Here, a frequency-constant loss factor $\eta_s=0.03$ is assumed as the typical loss factor for common building materials lies in an interval of $2\,\%$ to $5\,\%$ in the building-acoustics-relevant frequency range \cite{KLI2008}. $\eta_s$ is also displayed in Figure \ref{fig:eta}.\\
The volume velocity at the respective source positions in the source room is defined to be $Q_s=1e{-4}\,\text{m}^3/\text{s}$.\\
\\
The source and receiving room, as well as the gap, will be meshed using hexahedron fluid elements based on tri-quadratic Lagrangian shape functions. The walls' mesh consists of quadrilateral plane shell elements using bi-quadratic Lagrangian shape functions. While geometry and mesh creation are done in SALOME 9.14, the element formulations are implemented in elPaSo, i.e., the FE matrix assembly as well as solving takes place in elPaSo \cite{SRE2023}. \\
On the occasion of the  node set definition in SALOME, a comment on the positions of the sources and microphones should be given. In SALOME, node sets can be easily defined by grouping nodes of the mesh. Therefore, the source and microphone positions are chosen to be the closest node to the positions defined in Table \ref{tab:SourcePos}, Table \ref{tab:RecSRPos} and Table \ref{tab:RecRRPos}. As a consequence, the source and microphone positions change for different mesh sizes, which may result in a violation of the regulations defined in DIN EN ISO 10140-4 and DIN EN ISO 10140-5. Also, the comparison of the sound pressure at different microphone positions can be influenced due to the position shifts when comparing differently discretised meshes, which should be kept in mind during the mesh convergence study of the large-scale building acoustics test facility in Section \ref{Sec:Res_elPaSo_LSBATF}. \\
Now, based on the previously listed material data, the element lengths for the different meshes in the different domains can be determined by considering the wavelengths at the maximum frequency of interest - the procedure was already described in Subsection \ref{Sec:VMFEM_FSI}. The wavelength of the three materials listed above are shown over the building-acoustics-relevant frequency range in Figure \ref{fig:waveLengths}.
Evidently, glass wool modelled as an equivalent fluid always yields the smallest element length compared to air and plasterboard. It should be mentioned that the wavelength was calculated using the real part of the complex speed of sound in the respective domains; i.e. $\lambda=\text{Re}(\hat{c})/f$. \\
Nonetheless, before proceeding with the element length determination for the large-scale building acoustics test facility and the final estimations of the STLs, an intermediate step is taken by first verifying the in-house research code elPaSo using the commercial FE software COMSOL 6.3.

\section{elPaSo-Verification based on a Small-Scale Test Facility}\label{Sec:elPaSoVer}
The verification procedure will be conducted using a small-scale building acoustics test facility in the frequency range $1\,\text{Hz}$ to $1000\,\text{Hz}$. In doing so, an increasing complexity of the test facility is created by considering first a single-leaf wall as a test specimen (two coupling interfaces) and, afterwards, a double-leaf wall (four coupling interfaces). Since different loss factors $\eta_a$ will be applied to the air in the source and receiving room as well as the gap of the large-scale test facility, two single-leaf wall configurations (SLW1, SLW2) will be tested considering the loss factor, first, determined by the propagation loss $m$ ($\eta_{a,m}$) and second, by the reverberation time $T=1.5\,\text{s}$ ($\eta_{a,T}$). Then, two double-leaf wall configurations without and with insulation (DLWnI, DLWI) are tested, considering the loss factor determined by the reverberation time $T$ in the source and receiving room, while the loss factor of the gap-filling air is determined by the propagation loss $m$, if no insulation is applied. The four test cases are listed in Table \ref{tab:TestCasesVerif}.
\begin{table*}[!t]
    \caption{Test cases of the small-scale building acoustics test facility used during the verification of elPaSo. The loss factors in parentheses behind the materials denote the respective loss factor applied to the material, where $\eta_{a,m}$ and $\eta_{a,T}$ are the loss factors in air determined by the propagation loss $m$ and reverberation time $T=1.5\,\text{s}$, while $\eta_s$ is the loss factor in plasterboard.}\label{tab:TestCasesVerif}
    \vspace{0.3cm}
    \centering
        \begin{tabular}{lcccc}
        \hline
        Test Cases                                  & Source Room       & Gap               & Receiving Room        & Wall(s)  \\
        \hline
        Single-Leaf Wall 1 (SLW1)                   & Air ($\eta_{a,m}$)& -                 & Air ($\eta_{a,m}$)    & Plasterboard ($\eta_s$) \\
        Single-Leaf Wall 2 (SLW2)                   & Air ($\eta_{a,T}$)& -                 & Air ($\eta_{a,T}$)    & Plasterboard ($\eta_s$) \\
        \hline
        Double-Leaf Wall without\\ Insulation (DLWnI) & Air ($\eta_{a,T}$)& Air ($\eta_{a,m}$)& Air ($\eta_{a,T}$)    & Plasterboard ($\eta_s$) \\
        Double-Leaf Wall with\\ Insulation    (DLWI)  & Air ($\eta_{a,T}$)& Glass Wool        & Air ($\eta_{a,T}$)    & Plasterboard ($\eta_s$) \\
        \hline
        \end{tabular}
\end{table*}\\
The geometrical dimensions of the small-scale test facility are listed in Table \ref{tab:DimScBATFRooms}. 
\begin{table*}
    \caption{Dimensions of the small-scale building acoustics test facility's source and receiving room, as well as gap; while SLW resp. DLW are the acronyms for single-leaf resp. double-leaf wall configurations.}\label{tab:DimScBATFRooms}
    \vspace{0.3cm}
    \centering
        \begin{tabular}{lccc}
        \hline
                            & $l_x$ in m & $l_y$ in m & $l_z$ in m  \\
        \hline
        Source Room (SLW)   & $0.56$      & $1.05$      & $0.72$    \\
        Receiving Room (SLW)& $0.63$      & $1.05$      & $0.72$    \\
        \hline
        Source Room (DLW)   & $0.56$      & $1.05$      & $0.72$    \\
        Receiving Room (DLW)& $0.62$      & $1.05$      & $0.72$    \\
        Gap (DLW)           & $0.01$      & $1.05$      & $0.72$    \\
        \hline
        \end{tabular}
\end{table*}
In all test cases, two monopole sources excite the source room with a volume velocity $Q_s=1e{-4}\,\text{m}^3\text{/s}$. The respective sound source and microphone positions are listed in Table \ref{tab:SourceRecPosSCBATF}.
\begin{table*}[!t]
    \caption{Sound source positions $\text{S}_i$ in the source room as well as microphone positions $\text{SR}_i$ resp. $\text{RR}_i$ in the source resp. receiving room, $i=1,2$.}\label{tab:SourceRecPosSCBATF}
    \vspace{0.3cm}
    \centering
    \begin{tabular}{lcccccc}
    \hline
             & $\text{S}_1$ & $\text{S}_2$  & $\text{SR}_1$ & $\text{SR}_2$ & $\text{RR}_1$ & $\text{RR}_2$ \\
    \hline
    $x$ in m & $0.1$        & $0.3$         & $0.1$         & $0.2$         & $0.9$         & $0.6$         \\
    $y$ in m & $0.3$        & $0.8$         & $0.4$         & $0.8$         & $0.6$         & $0.2$         \\
    $z$ in m & $0.1$        & $0.4$         & $0.7$         & $0.2$         & $0.6$         & $0.4$         \\
    \hline
    \end{tabular}
\end{table*}\\
Now, referring to Figure \ref{fig:waveLengths}, the minimal element lengths will be determined for the meshes used in elPaSo and COMSOL - both meshes will consist of hexahedron fluid and quadrilateral shell elements with quadratic shape functions. In the case of the single-leaf wall test specimens (SLW1, SLW2) as well as the double-leaf wall test specimen without insulation (DLWnI), the sound waves are resolved with 12 nodes per wavelength; 7 nodes per wavelength are used for the double-leaf wall test specimen with insulation (DLWI). According to the rule of thumb, 12 nodes per wavelength provides a finer mesh than necessary, while 7 nodes per wavelength provides a coarser mesh. The insufficient sampling rate is reasoned in the computational effort; however, at this point in the verification process, it was already expected that elPaSo and COMSOL should yield the same results regardless of the sampling rate. Now, the element lengths for 7 resp. 12 nodes per wavelength are for the fluid $l_{e,a}\approx0.11\,\text{m}$ resp. $l_{e,a}\approx0.06\,\text{m}$, the equivalent fluid $l_{e,e}\approx0.03\,\text{m}$, and the structure $l_{e,s}\approx0.09\,\text{m}$ resp. $l_{e,s}\approx0.05\,\text{m}$. Consequently, the single-leaf wall test specimens and double-leaf wall without insulation will be meshed using $l_{e,s}\approx0.05\,\text{m}$, while the mesh of the double-leaf wall with insulation is based on $l_{e,e}\approx0.03\,\text{m}$. Then, the mesh of the single-leaf wall test specimens comprises $n_\text{SLW1}=n_\text{SLW2}=77315$ degrees of freedom, and the double-leaf wall without insulation $n_\text{DLWnI}=89311$ resp. with insulation $n_\text{DLWI}=337463$ degrees of freedom.\\
Lastly, it should also be mentioned that the complex speed of sound due to damping was calculated before the assembly of the matrix in elPaSo and COMSOL based on the loss factors in Equation (\ref{eqn:LossFactorsAir}). Then, the same data set of the complex speed of sound over the frequency was processed to elPaSo and COMSOL in order to ensure that the same material properties underlie the subsequent computations. The same applies for the glass wool's complex speed of sound and effective density. \\
\\
During the verification process, the sound pressure $p_j$ at the microphone position $j$, $j=1,2,3,4$, obtained from elPaSo was compared with corresponding results from COMSOL using the relative error 
\begin{equation}
    \epsilon_{r_{i,j}}(f_i)=\frac{p_{j,\text{elPaSo}}(f_i)-p_{j,\text{COMSOL}}(f_i)}{p_{j,\text{elPaSo}}(f_i)}
\end{equation}
at each computed frequency $f_i$, $i=1,2,...,1000$, (step size $\Delta f=1\,\text{Hz}$) as well as the averaged relative error over the considered frequency range
\begin{equation}
    \epsilon_{r_j,ave} = \frac{1}{1000}\sum_{i=1}^{1000} \epsilon_{r_{i,j}}(f_i).
\end{equation}
Unfortunately, COMSOL was unable to mesh the small-scale test facility with a double-leaf test specimen and insulation as long as the microphone position $\text{SR}_1$ was included due to mesh distortion. Thus, this receiver position is neglected for this specific configuration. For reasons of scope, two representative microphone positions, $\text{SR}_2$ and $\text{RR}_2$, will be presented subsequently. 

\subsection{Single-Leaf Wall}
Beginning with the single-leaf wall test specimen and atmospheric damping in air (SLW1), Figure \ref{fig:SLW_dampAtmos_SR2} and Figure \ref{fig:SLW_dampAtmos_RR2} display the sound pressure levels (SPL) over the frequency range at the two representative microphone positions.
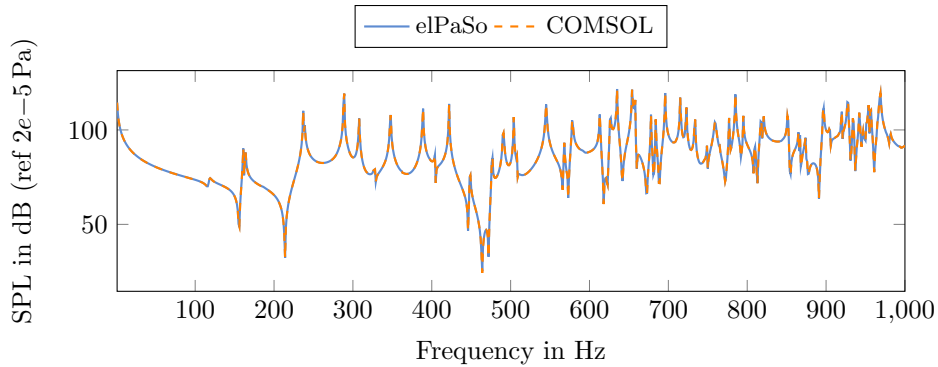
\begin{figure}[b!]
    \centering
        \begin{tikzpicture}
            \begin{axis}[
                xlabel={Frequency in Hz},
                ylabel={SPL in dB (ref $2e{-5}\,\text{Pa}$)},
                xmin=1, xmax=1000,
                width=12cm,
                height=4.5cm,
                scaled y ticks=true,
                legend style={
                at={(0.5,1.2)},
                anchor=center,
                draw=black,
                legend cell align=left,
                fill=none,
                legend columns=3,
                font=\small
                }
                ]
                \addplot [mark=none, color=spec_blue, thick] table [
                    x index= 0, 
                    y index= 1 
                    ] {meanSolution_elPaSo_AirAtmospheric_77k_SR2.txt};
                \addlegendentry{elPaSo}
                \addplot [mark=none, color=orange, thick, dashed] table [
                    x index=0, 
                    y index=1
                    ] {meanSolution_soundPressure_SR2_1_1000Hz_dampedAtmospheric.txt};
                \addlegendentry{COMSOL}
            \end{axis}
        \end{tikzpicture}
    \caption{Single-leaf wall configuration (SLW1): sound pressure level (SPL) at microphone position $\text{SR}_2$ computed using elPaSo and COMSOL with an maximum relative error $\epsilon_{r,max}=0.11$ and average relative error $\epsilon_{r,ave}=0.0$.}
    \label{fig:SLW_dampAtmos_SR2}
\end{figure}
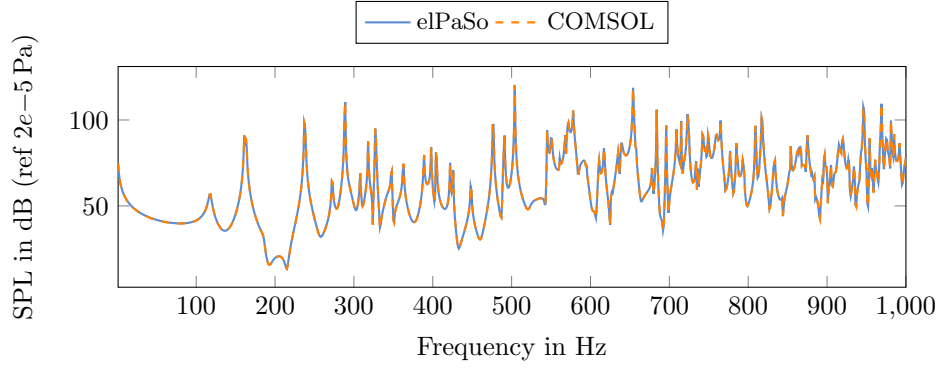
\begin{figure}[t!]
    \centering
        \begin{tikzpicture}
            \begin{axis}[
                xlabel={Frequency in Hz},
                ylabel={SPL in dB (ref $2e{-5}\,\text{Pa}$)},
                xmin=1, xmax=1000,
                width=12cm,
                height=4.5cm,
                scaled y ticks=true,
                legend style={
                at={(0.5,1.2)},
                anchor=center,
                draw=black,
                legend cell align=left,
                fill=none,
                legend columns=3,
                font=\small
                }
                ]
                \addplot [mark=none, color=spec_blue, thick] table [
                    x index= 0, 
                    y index= 1 
                    ] {meanSolution_elPaSo_AirAtmospheric_77k_RR2.txt};
                \addlegendentry{elPaSo}
                \addplot [mark=none, color=orange, thick, dashed] table [
                    x index=0, 
                    y index=1
                    ] {meanSolution_soundPressure_RR2_1_1000Hz_dampedAtmospheric.txt};
                \addlegendentry{COMSOL}
            \end{axis}
        \end{tikzpicture}
    \caption{Single-leaf wall configuration (SLW1): sound pressure level (SPL) at microphone position $\text{RR}_2$ computed using elPaSo and COMSOL, with an maximum relative error $\epsilon_{r,max}=0.04$ and average relative error $\epsilon_{r,ave}=0.01$.}
    \label{fig:SLW_dampAtmos_RR2}
\end{figure}
A visual consistency between elPaSo's and COMSOL's SPL exists, which is highlighted by the average relative error over the frequency that is $e_{r,ave} =0.0$ resp. $e_{r,ave}=0.01$ at $\text{SR}_2$ resp. $\text{RR}_2$. The maximum relative error at $\text{SR}_2$ resp. $\text{RR}_2$ is $\epsilon_{r,max}=0.11$ resp. $\epsilon_{r,max}=0.04$.\\
Similar results yields the single-leaf wall configuration SLW2, where the air in the source and receiving room is damped to a reverberation time $T=1.5\,\text{s}$. The respective SPL plots are shown in Figure \ref{fig:SLW_dampTrev_SR2} and Figure \ref{fig:SLW_dampTrev_RR2} (Appendix \ref{App:VerifEl}). Compared to test case SLW1, the average relative errors are the same, while the maximum relative error vary for both test cases at the microphone positions. In conclusion, elPaSo is able to yield a sufficiently accurate solution compared to COMSOL for a single-leaf test specimen assuming different damping types in the source and receiving room's air. Hence, the verification can be further conducted for the double-leaf wall test specimen with and without insulation.

\subsection{Double-Leaf Wall}
First, the double-leaf wall test specimen without insulation (DLWnI) is simulated, where the air is damped to the reverberation time ($\eta_{a,T}$) in the source and receiving room as well as with respect to atmospheric damping ($\eta_{a,m}$) in the gap. Now, Figure \ref{fig:DLW_noInsu_SR2} and Figure \ref{fig:DLW_noInsu_RR2} show the SPL for the two representative microphone positions from which, again, a visual consistency arises.
\begin{figure}[b!]
    \centering
        \begin{tikzpicture}
            \begin{axis}[
                xlabel={Frequency in Hz},
                ylabel={SPL in dB (ref $2e{-5}\,\text{Pa}$)},
                xmin=1, xmax=1000,
                width=12cm,
                height=4.5cm,
                scaled y ticks=true,
                legend style={
                at={(0.5,1.2)},
                anchor=center,
                draw=black,
                legend cell align=left,
                fill=none,
                legend columns=3,
                font=\small
                }
                ]
                \addplot [mark=none, color=spec_blue, thick] table [
                    x index= 0, 
                    y index= 1 
                    ] {meanSolution_elPaSo_noInsulation_89k_SR2.txt};
                \addlegendentry{elPaSo}
                \addplot [mark=none, color=orange, thick, dashed] table [
                    x index=0, 
                    y index=1
                    ] {meanSolution_soundPressure_SR2_1_1000Hz_noInsulation.txt};
                \addlegendentry{COMSOL}
            \end{axis}
        \end{tikzpicture}
    \caption{Double-leaf wall configuration without insulation (DLWnI): sound pressure level (SPL) at microphone position $\text{SR}_2$ computed using elPaSo and COMSOL, with an maximum relative error $\epsilon_{r,max}=0.01$ and average relative error $\epsilon_{r,ave}=0.0$.}
    \label{fig:DLW_noInsu_SR2}
\end{figure}
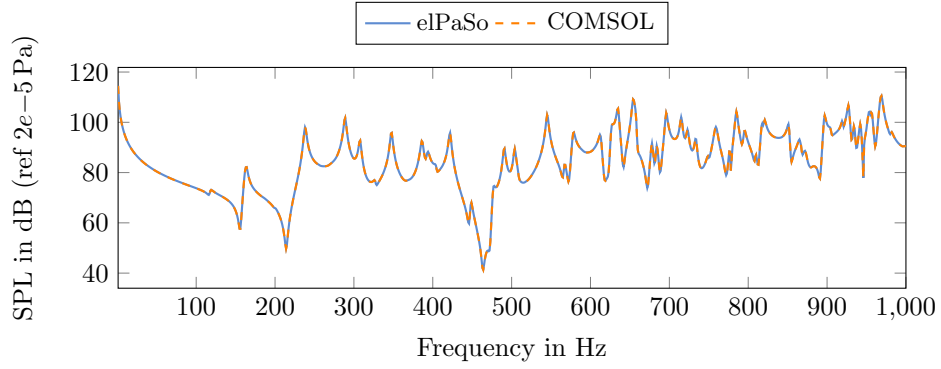
\begin{figure}[b!]
    \centering
        \begin{tikzpicture}
            \begin{axis}[
                xlabel={Frequency in Hz},
                ylabel={SPL in dB (ref $2e{-5}\,\text{Pa}$)},
                xmin=1, xmax=1000,
                width=12cm,
                height=4.5cm,
                scaled y ticks=true,
                legend style={
                at={(0.5,1.2)},
                anchor=center,
                draw=black,
                legend cell align=left,
                fill=none,
                legend columns=3,
                font=\small
                }
                ]
                \addplot [mark=none, color=spec_blue, thick] table [
                    x index= 0, 
                    y index= 1 
                    ] {meanSolution_elPaSo_noInsulation_89k_RR2.txt};
                \addlegendentry{elPaSo}
                \addplot [mark=none, color=orange, thick, dashed] table [
                    x index=0, 
                    y index=1
                    ] {meanSolution_soundPressure_RR2_1_1000Hz_noInsulation.txt};
                \addlegendentry{COMSOL}
            \end{axis}
        \end{tikzpicture}
    \caption{Double-leaf wall configuration without insulation (DLWnI): sound pressure level (SPL) at microphone position $\text{RR}_2$ computed using elPaSo and COMSOL, with an maximum relative error $\epsilon_{r,max}=0.07$ and average relative error $\epsilon_{r,ave}=0.01$.}
    \label{fig:DLW_noInsu_RR2}
\end{figure}
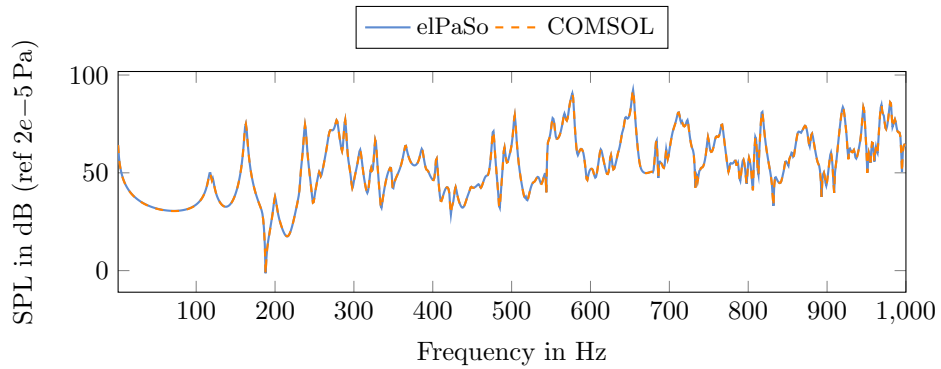
This is supported considering the average relative error $e_{r,ave}=0.0$ resp. $e_{r,ave}=0.01$ at $\text{SR}_2$ resp. $\text{RR}_2$, while the maximum relative error is $e_{r,max}=0.01$ resp. $e_{r,max}=0.07$. Next to this, the sound pressure yielding the SPL in Figure \ref{fig:DLW_Insu_SR2} and Figure \ref{fig:DLW_Insu_RR2} (Appendix \ref{App:VerifEl}) for the double-leaf wall test specimen with insulation (DLWI) results in the same averaged error compared to DLWnI, while the maximum relative error differs. However, it is again valid to draw the conclusion that elPaSo is able to compute an sufficiently accurate solution compared to COMSOL. \\
\\
Overall, the element and coupling formulations implemented in elPaSo are verified using COMSOL, which then offers the possibility to exclusively model the large-scale building acoustics test facility using SALOME and elPaSo. The reason COMSOL is not further used as a comparison is that it is not straightforward to create non-conforming meshes at an interface combining three- and two-dimensional elements in COMSOL 6.3. Instead, conforming meshes must be used, which are determined by the smallest element length. In the case presented here, the smallest element length is always given by the equivalent fluid as already mentioned at the end of Section \ref{Sec:BATF_FEM}. When also demanding a sufficiently small element length that yields physical results, creating a conforming mesh for the large-scale building acoustics test facility in the considered frequency range would yield a time and resource exhaustive meshing procedure and STL estimation due to the size of the resulting linear system of equations.

\section{elPaSo-Model of a Large-Scale Test Facility}\label{Sec:Res_elPaSo_LSBATF}
The subsequent section is structured as follows: first, in Subsection \ref{Sec:Res_MeshConv}, a mesh convergence study is conducted to identify the optimal number of nodes per wavelength in order to determine the minimum required element length in the respective domains. Second, in Subsection \ref{Sec:Res_SLW}, the STL of a single-leaf wall test specimen is computed, where the mass is equal to the total mass of the two leafs of the double-leaf wall test specimen computed in Subsection \ref{Sec:Res_DLW}. Compared to the STL definition in Equation (\ref{eqn:STL}), the correction term $10\text{lg}(S/A)$ will be neglected in the following owing to the well-defined conditions in the receiving room. The large-scale building acoustics test facility described in Section \ref{Sec:BATF_FEM} is used for all test specimen cases. Nevertheless, the single-leaf wall test specimen was not specified, which is simply a plasterboard wall of thickness $h=0.05\,\text{m}$. And, compared to the double-leaf wall test specimen, the gap between the two leafs is not existent in the case of the single-leaf wall test specimen. This also means that the offset in Table \ref{tab:RecRRPos} is now only $l_{x_\text{SR}}=5\,\text{m}$.

\subsection{Mesh Convergence Study}\label{Sec:Res_MeshConv}
Despite the fact that a single-leaf wall as well as an double-leaf test specimen will be considered, the mesh convergence study is conducted only for the double-leaf wall with and without insulation. It is to be assumed that if a mesh convergence of the double-leaf wall is reached, this also applies to the less complex case of the single-leaf wall. Furthermore, the mesh convergence study will be conducted in the frequency range $100\,\text{Hz}$ to $300\,\text{Hz}$. \\
The mesh convergence study consists of the comparison of four different sampling strategies: $7$, $10$, $13$ and $16$ nodes per wavelength. According to the rule of thumb, $7$ nodes per wavelength is too coarse, while the remaining sampling rates are at least sufficient. The resulting element length and degrees of freedom according to the wavelengths in Figure \ref{fig:waveLengths} are listed in Table \ref{tab:ElemSizesConvStud}.
\begin{table*}
    \caption{Mesh convergence study of the large-scale building acoustics test facility with a double-leaf wall with and without insulation (DLWI, DLWnI) in the frequency range $100\,\text{Hz}$ to $300\,\text{Hz}$: sampling rates (nodes per wavelength) for the discretisation, resulting element lengths and degrees of freedom of the FE Models.}
    \vspace{0.3cm}
    \label{tab:ElemSizesConvStud}
    \centering
    \begin{tabular}{cccccc}
        \hline
        \multirow{2}{*}{Nodes per Wavelength}   & \multicolumn{3}{c}{Element Length in m} & \multicolumn{2}{c}{Degrees of Freedom}\\
        \cline{2-6}
                                                & Structure & Fluid & Equivalent Fluid  & DLWnI     & DLWI                       \\
        \hline
        $\sim 7$                                & $0.18$    & $0.38$& $0.06$            & $42027$   & $81759$                   \\
        $\sim 10$                               & $0.12$    & $0.25$& $0.04$            & $109053$  & $258333$                  \\
        $\sim 13$                               & $0.09$    & $0.19$& $0.03$            & $234243$  & $500133$                  \\
        $\sim 16$                               & $0.07$    & $0.15$& $0.02$            & $422063$  & $1260205$                 \\
        \hline
    \end{tabular}
\end{table*}
The sampling rates are denoted as the approximate target values since the element lengths are rounded. \\
In order to yield a statement about the convergence, the mean sound pressure at all receiver positions in the source resp. receiving room will be compared in the sense of the
frequency response assurance criterion (FRAC) \cite{BLE2017}:
\begin{equation}\label{eqn:FRAC}
 FRAC(j) = \frac{\left(H_1(f,j)^T\cdot H_2(f,j)\right)^2}{\left(H_1(f,j)^T\cdot H_1(f,j)\right)\left(H_2(f,j)^T\cdot H_2(f,j)\right)},
\end{equation}
where $H_i(f,j)$, $i=1,2$, are the frequency response functions of the $j$-th degree of freedom. Here, the frequency response function of the $j$-th degree of freedom is substituted with the mean sound pressure over the eight respective microphone positions in the source resp. receiving room for a certain sampling rate; i.e. $7$ nodes per wavelength is compared to $10$, $10$ to $13$, and $13$ to $16$. So, technically it is not the FRAC in the sense of the definition in Equation (\ref{eqn:FRAC}); nevertheless, for the sake of simplicity, it will be referred to as FRAC in the following. Moreover, the FRAC is not determined over the entire frequency range, but in $25\,\text{Hz}$-intervals from $100\,\text{Hz}$ to $300\,\text{Hz}$. \\
In the following, only the results of the mesh convergence study of the double-leaf wall with insulation are shown due to the reason of scope. As can be seen in Figure \ref{fig:SPL_7_16nodes_SR} (Appendix \ref{App:BATF_Res}) and Figure \ref{fig:SPL_7_16nodes_RR}, the mean sound pressure level at the microphone positions in the source and receiving room converge, which is also mathematically supported by the FRACs in Figure \ref{fig:FRAC_SR} (Appendix \ref{App:BATF_Res}) and Figure \ref{fig:FRAC_RR}.
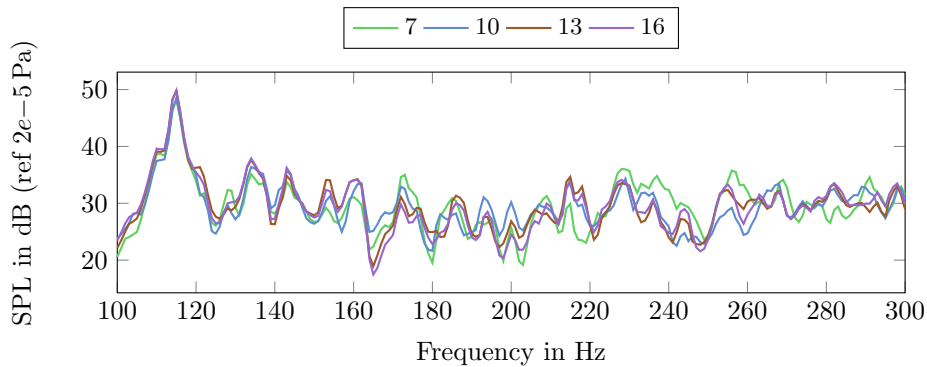
\begin{figure}[b!]
    \centering
        \begin{tikzpicture}
            \begin{axis}[
                ylabel={SPL in dB (ref $2e{-5}\,\text{Pa}$)},
                xlabel={Frequency in Hz},
                xmin=100, xmax=300,
                width=12cm,
                height=4.5cm,
                scaled y ticks=true,
                legend style={
                at={(0.5,1.2)},
                anchor=center,
                draw=black,
                legend cell align=left,
                fill=none,
                legend columns=4,
                font=\small
                }
                ]
                \addplot [color=spec_green, thick] table [
                    x index=0, 
                    y index=1
                    ] {ConvStud_7nodes_Insulation_RR.txt};
                \addlegendentry{7}
                \addplot [color=spec_blue, thick] table [
                    x index= 0, 
                    y index= 1 
                    ] {ConvStud_10nodes_Insulation_RR.txt};
                \addlegendentry{10}
                \addplot [color=spec_brown, thick] table [
                    x index= 0, 
                    y index= 1 
                    ] {ConvStud_13nodes_Insulation_RR.txt};
                \addlegendentry{13}
                \addplot [color=spec_amethyst, thick] table [
                    x index= 0, 
                    y index= 1 
                    ] {ConvStud_16nodes_Insulation_RR.txt};
                \addlegendentry{16}
            \end{axis}
        \end{tikzpicture}
    \caption{Mean sound pressure level (SPL) over all receiver positions in the receiving room of the large-scale building acoustics test facility in the frequency range $100\,\text{Hz}$ to $300\,\text{Hz}$ based on meshes with quadratic elements utilising element lengths according to 7, 10, 13 and 16 nodes per wavelength.}
    \label{fig:SPL_7_16nodes_RR}
\end{figure}
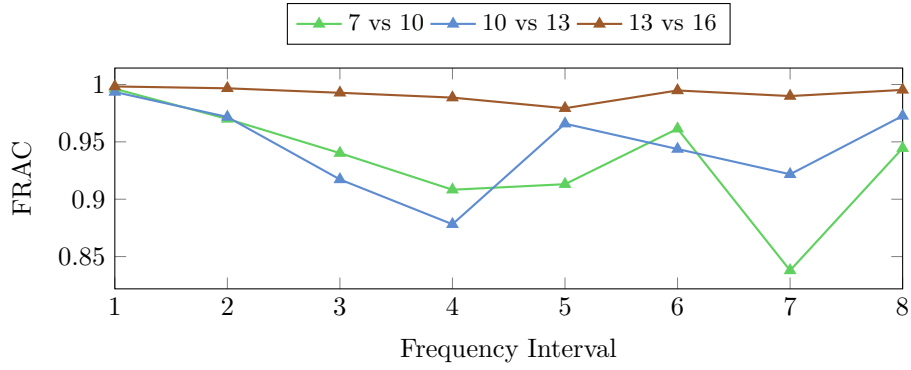
\begin{figure}%[t!]
    \centering
        \begin{tikzpicture}
            \begin{axis}[
                ylabel={FRAC},
                xlabel={Frequency Interval},
                xmin=1, xmax=8,
                width=12cm,
                height=4.5cm,
                scaled y ticks=true,
                legend style={
                at={(0.5,1.2)},
                anchor=center,
                draw=black,
                legend cell align=left,
                fill=none,
                legend columns=3,
                font=\small
                }
                ]
                \addplot [mark=triangle*, color=spec_green, thick] table [
                    x index=0, 
                    y index=1
                    ] {Frac_7_10nodes_Insulation_RR.txt};
                \addlegendentry{7 vs 10}
                \addplot [mark=triangle*, color=spec_blue, thick] table [
                    x index= 0, 
                    y index= 1 
                    ] {Frac_10_13nodes_Insulation_RR.txt};
                \addlegendentry{10 vs 13}
                \addplot [mark=triangle*, color=spec_brown, thick] table [
                    x index= 0, 
                    y index= 1 
                    ] {Frac_13_16nodes_Insulation_RR.txt};
                \addlegendentry{13 vs 16}
            \end{axis}
        \end{tikzpicture}
    \caption{Mesh convergence study of the large-scale building acoustics test facility with a double-leaf wall and insulation utilising the FRAC criterion in $25\,\text{Hz}$ intervals in the frequency range $100\,\text{Hz}$ to $300\,\text{Hz}$. Here, the mean sound pressure at all receiver positions in the receiving room are compared arising from a mesh with quadratic elements where the element length is determined by samplings with 7, 10, 13 and 16 nodes per wavelength..}
    \label{fig:FRAC_RR}
\end{figure}
However, the FRACs of the receiving room are smaller when comparing the sampling rates $7$ to $10$ nodes and $10$ to $13$ nodes - i.e. the mesh convergence is slower at the receiver positions in the receiving room compared to those in the source room. Hence, the mesh convergence has to be determined by the FRACs of the receiving room in Figure \ref{fig:FRAC_RR}. While comparing the sampling rates $7$ to $10$ and $10$ to $13$ nodes per wavelength, the FRACs in the intervals $1$, $2$, $3$, $4$ and $6$ are close to each other, indicating that $7$ nodes per wavelength could be enough. Nonetheless, the FRACs of comparing $13$ to $16$ nodes per wavelength reveal in all frequency intervals a value close to $1$, while the FRACs of the sampling rate comparisons $7$ to $10$ and $10$ to $13$ are mostly not close to $1$. Consequently, $13$ nodes per wavelength is the best sampling rate according to the mesh convergence study in the frequency range $100\,\text{Hz}$ to $300\,\text{Hz}$. \\
\\
Based on this result, the required element lengths can be determined for the single-leaf and double-leaf wall test specimens with and without insulation. Since the aim of this contribution is to provide a comprehensive data set for the simulation of a simple building acoustics test facility, the size of the meshes shall be kept in a range that the solving process can be handled in a decent time by commercially available workstations. So, the following model sizes do not indicate the limits of elPaSo. \\
Here, the STL will be determined in the frequency range $1\,\text{Hz}$ to $715\,\text{Hz}$ - $1\,\text{Hz}$ to $99\,\text{Hz}$ is also considered as the double-leaf wall resonance is smaller than $100\,\text{Hz}$. Based on this, a frequency- and domain-specific discretisation strategy, cf. \cite{BLE2024}, is used. Domain-specific as non-conforming meshes will be constructed using a different discretisation in every domain with respect to the minimal required element length; and frequency-specific since the element lengths are chosen so that they can only resolve wave phenomena up to a certain maximum frequency. This means instead of using a mesh with minimum element lengths required for resolving vibroacoustic wave phenomena up to $715\,\text{Hz}$, four different meshes will be created resolving vibroacoustic wave phenomena up to $300\,\text{Hz}$, $500\,\text{Hz}$, $650\,\text{Hz}$ and $715\,\text{Hz}$, which results in the meshes documented in Table \ref{tab:ElemSizes13nodes}.
\begin{table*}[!b]
    \caption{Frequency- and domain-specific meshes of the large-scale building acoustics test facility with a single-leaf (SLW) and double-leaf wall with and without insulation (DLWI, DLWnI) using $\sim13$ nodes per wavelength in the frequency range $1\,\text{Hz}$ to $715\,\text{Hz}$: resulting element lengths and degrees of freedom.}
    \vspace{0.3cm}
    \label{tab:ElemSizes13nodes}
    \centering
    \begin{tabular}{ccccccc}
        \hline
        \multirow{2}{*}{Frequency in Hz }       & \multicolumn{3}{c}{Element Length in m} & \multicolumn{3}{c}{Degrees of Freedom}        \\
        \cline{2-7}
                                                & Structure & Fluid & Equivalent Fluid  & SLW           & DLWnI         & DLWI          \\
        \hline
        $\left[1,300\right]$                    & $0.09$    & $0.19$& $0.03$            & $192114$      & $234243$      & $500133$      \\
        $\left[300,500\right]$                  & $0.07$    & $0.11$& $0.02$            & $813474$      & $887373$      & $1719455$     \\
        $\left[500,650\right]$                  & $0.06$    & $0.09$& $0.02$            & $1425516$     & $1526163$     & $2352233$     \\
        $\left[650,715\right]$                  & $0.057$   & $0.08$& $0.019$           & $1973840$     & $2088977$     & $3004283$     \\
        \hline
    \end{tabular}
\end{table*}\\
The advantage of the domain- and frequency-specific discretisation approach becomes evident when investigating the solution times listed in Table \ref{tab:CompTimes13Nodes}.
\begin{table*}
    \caption{Solution times and maximum used RAM for the different frequency- and domain-specific meshes of the large-scale building acoustics test facility with a single-leaf (SLW) and double-leaf wall with and without insulation (DLWI, DLWnI). The computations were undertaken on a work station with two Intel\textsuperscript\textregistered Xeon\textsuperscript\textregistered Gold $6138$ @ $2.00\,\text{GHz}$ processors, $40$ cores ($80$ threads with hyper-threading), $755\,\text{GB}$ RAM and Ubuntu 22.04.3 LTS as the operating system. elPaSo was called using 80 threads (4 parallel processes, each with 20 threads) and MUMPS for solving the linear system of equations.}
    \vspace{0.3cm}
    \label{tab:CompTimes13Nodes}
    \centering
    \begin{tabular}{ccccccccccc}
        \hline
        \multirow{2}{*}{\makecell{Frequency\\in Hz}}   & \multicolumn{3}{c}{Degrees of Freedom}        & \multirow{2}{*}{Steps}& \multicolumn{3}{c}{Solution Time/Step in s}   & \multicolumn{3}{c}{max. RAM in GB}\\
        \cline{2-4}\cline{6-11}
                                            & SLW           & DLWnI         & DLWI          &                       & SLW       & DLWnI     & DLWI                  & SLW   & DLWnI & DLWI              \\
        \hline
        $\left[1,300\right]$                & $192114$      & $234243$      & $500133$      & $300$                 & $5.3$     & $8$       & $25$                  & $3.95$& $5.0$ & $11.9$            \\
        $\left[300,500\right]$              & $813474$      & $887373$      & $1719455$     & $201$                 & $37$      & $64$      & $110$                 & $17.5$& $18.7$& $37.9$            \\
        $\left[500,650\right]$              & $1425516$     & $1526163$     & $2352233$     & $151$                 & $161$     & $148$     & $202$                 & $30.8$& $32.4$& $49.1$            \\
        $\left[650,715\right]$              & $1973840$     & $2088977$     & $3004283$     & $66$                  & $254$     & $266$     & $370$                 & $41.6$& $42.0$& $62.5$            \\
        \hline\hline
        $\left[1,715\right]$                &               &               &               & $718$                 & $50102$   &$55168$    &$84532$               &       &       &                   \\
        \hline
    \end{tabular}
\end{table*}
elPaSo is used on a workstation with two Intel\textsuperscript\textregistered Xeon\textsuperscript\textregistered Gold $6138$ @ $2.00\,\text{GHz}$ processors, $40$ cores ($80$ threads with hyper-threading), $755\,\text{GB}$ RAM and Ubuntu 22.04.3 LTS as the operating system. In each simulation run, $80$ threads were running: 4 parallel processes, each with 20 threads. MUMPS (MUltifrontal Massively Parallel sparse direct Solver) was used for solving the linear system of equations. The total solution times for the single-leaf wall, double-leaf wall without insulation and double-leaf wall with insulation are $50102\,\text{s}\approx13.9\,\text{h}$, $55168\,\text{s}\approx15.3\,\text{h}$ and $84532\,\text{s}\approx23.5\,\text{h}$. In contrast, using only a domain-specific discretisation and the meshes capable of resolving vibroacoustic wave phenomena up to $715\,\text{Hz}$ ($715$ steps), the total solution times for the single-leaf wall, double-leaf wall without insulation and double-leaf wall with insulation would be $181610\,\text{s}\approx50.4\,\text{h}$, $190190\,\text{s}\approx52.8\,\text{h}$ and $264550\,\text{s}\approx73.5\,\text{h}$; i.e., $3.6$-, $3.4$- and $3.1$-times as much. It should be mentioned that the solution times are determined by a single simulation run. To yield more representative run times, several simulation runs should be conducted - however, specific efficiency documentation are not the main objective of this contribution. Also, Table \ref{tab:CompTimes13Nodes} shows the maximum RAM used during the simulation runs, which, again, underlines that the solving process can be handled by commercially available workstations - here only $8\,\%$ of the total RAM was used. Nonetheless, in future works, a more meaningful way would be to parallelise also the solution of not only one but multiple frequencies instead of concentrate all $80$ threads on solving one frequency step. Next to this, further efficiency gains, and therefore possibilities to extend the investigated frequency range, are the use of efficient solving strategies, high-performance computing (HPC) clusters - e.g., the phoenix cluster at TU Braunschweig - as well as surrogate models, for instance, obtained by model order reduction (MOR), see \cite{BLE2024}, \cite{ROE2021} and \cite{SRE2021_1}. \\
With the computed sound pressure fields in the source and receiving room at hand, first, the STL for a single-leaf wall can be determined, and second for the double-leaf wall with and without insulation.

\subsection{Single-Leaf Wall}\label{Sec:Res_SLW}
\begin{figure}[b!]
    \centering
        \begin{tikzpicture}
            \begin{semilogxaxis}[
                xlabel={Frequency in Hz},
                ylabel={$R$ in dB (ref $2e{-5}\,\text{Pa}$)},
                xmin=10, xmax=630,
                ymin=0, ymax=60,
                xtick={16,31.5,63,125,250,500},
                xticklabels={16,31.5,63,125,250,500},
                width=12cm,
                height=4.5cm,
                scaled y ticks=true,
                legend style={
                at={(0.5,1.3)},
                anchor=center,
                draw=black,
                legend cell align=left,
                fill=none,
                legend columns=2,
                font=\small
                }
                ]
                \addplot [mark=triangle*,color=spec_blue, thick] table [
                    x index=0, 
                    y index=1
                    ] {STLthO_SLW.txt};
                \addlegendentry{STL}
                \addplot [densely dotted,color=spec_orange, thick] table [
                    x index= 0, 
                    y index= 1 
                    ] {massImpedanceLaw_SLW.txt};
                \addlegendentry{Mass Impedance}
                \addplot [dashed,color=spec_green, thick] coordinates {(14.41,0) (14.41,70)};
                \addplot [dashed,color=spec_green, thick] coordinates {(33.91,0) (33.91,70)};
                \addplot [dashed,color=spec_green, thick] coordinates {(63.53,0) (63.53,70)};
                \addplot [dashed,color=spec_green, thick] coordinates {(24.19,0) (24.19,70)};
                \addplot [dashed,color=spec_green, thick] coordinates {(42.97,0) (42.97,70)};
                \addplot [dashed,color=spec_green, thick] coordinates {(72.34,0) (72.34,70)};
                \addplot [dashed,color=spec_green, thick] coordinates {(40.30,0) (40.30,70)};
                \addplot [dashed,color=spec_green, thick] coordinates {(58.19,0) (58.19,70)};
                \addplot [dashed,color=spec_green, thick] coordinates {(86.95,0) (86.95,70)};
                \addlegendentry{Resonances Single-Leaf Wall}
            \end{semilogxaxis}
        \end{tikzpicture}
    \caption{Sound transmission loss (STL) of the large-scale building acoustics test facility with a single-leaf wall test specimen in one-third-octave bands over the frequency range $10\,\text{Hz}$ to $630\,\text{Hz}$. The largest displayed single-leaf wall resonance is $f_0=86.95\,\text{Hz}$.}
    \label{fig:STL_SLW}
\end{figure}
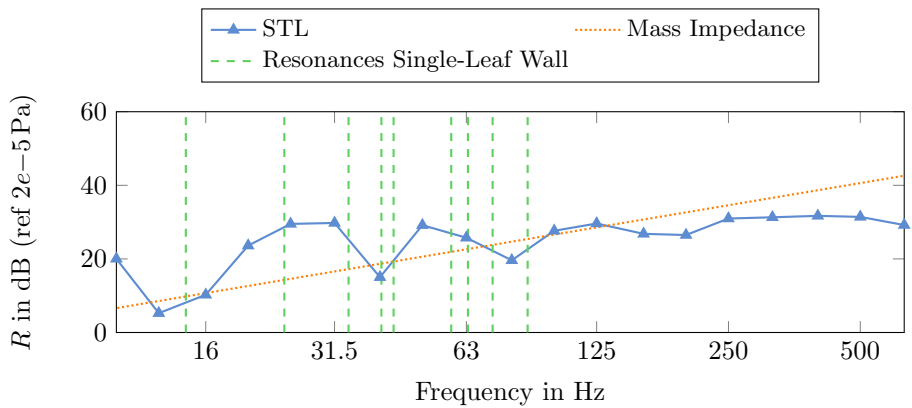

The numerically derived STLs for the single-leaf wall, and later also for the double-leaf wall, will be compared to theoretical STL profiles that can be found in literature, e.g. \cite{SIN2020}, \cite{VIG2008} or \cite{WIL2020}. For the single-leaf wall, the STL profile can be divided into three zones: first, the STL is determined by Berger's mass impedance law; second, the area of decline due to the coincidence frequency; and third, an improvement of the STL compared to the prediction of Berger's mass impedance law. Berger's mass impedance law for an oblique sound incidence averaged over the incidence angle $0^{\circ}\leq\varphi\leq 90^{\circ}$ reads \cite{SIN2020}
\begin{equation}\label{eqn:Berger}
    R_f=20\text{lg}\left(\frac{\pi f m''}{Z}\right)\,\text{dB}-3\,\text{dB},
\end{equation}
with the areal mass $m''=\rho h$ and the impedance of air $Z=\rho_0 c_0$. The STL measured in a diffuse sound field is reduced by $3\,\text{dB}$ compared to vertical sound incidence. The coincidence frequency of an infinitely large plate based on Kirchhoff's plate theory is \cite{SIN2020}
\begin{equation}\label{eqn:Coincide}
    f_c=\frac{c^2}{2\pi}\sqrt{\frac{m''}{B}}.
\end{equation}
As the Mindlin theory underlies the plane shell elements in elPaSo, it is to be expected that the decline due to the coincidence frequency in the following STL plots does not exactly match the Kirchhoff-theory-based prediction in Equation (\ref{eqn:Coincide}), since Kirchhoff neglects shear effects in plate bending. For very small frequencies, the STL profile is determined by declines caused by the eigenfrequencies of the wall. Here, the eigenfrequencies of a fully clamped Kirchhoff plate are used according to \cite{LEI1969}. Again, this is only an estimation due to the difference between the Kirchhoff and Mindlin theory; it is not to be expected that the declines in the numerically determined STL profiles match exactly the predicted eigenfrequencies. \\
Figure \ref{fig:STL_SLW} displays the STL averaged in one-third octave bands from $8\,\text{Hz}$ to $630\,\text{Hz}$. 
The first nine eigenfrequencies of the Kirchhoff plate as well as Berger's mass impedance law are also shown. It can be observed, that the STL profile reveals several declines in the frequency range of the first nine eigenfrequencies. Following this, an increase according to Berger's mass impedance law cannot be properly observed, but the coincidence frequency is already at $f_0\approx662\,\text{Hz}$, which could explain the beginning decline at $630\,\text{Hz}$. However, to gain meaningful insights on the single-leaf wall's STL profile, the frequency range needs to be expanded towards higher frequencies. 

\subsection{Double-Leaf Wall}\label{Sec:Res_DLW}
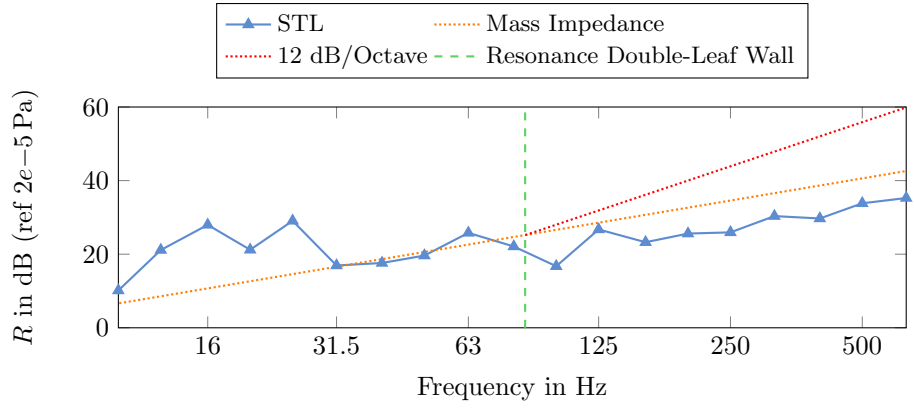
\begin{figure}[!t]
    \centering
        \begin{tikzpicture}
            \begin{semilogxaxis}[
                xlabel={Frequency in Hz},
                ylabel={$R$ in dB (ref $2e{-5}\,\text{Pa}$)},
                xmin=10, xmax=630,
                ymin=0, ymax=60,
                xtick={16,31.5,63,125,250,500},
                xticklabels={16,31.5,63,125,250,500},
                width=12cm,
                height=4.5cm,
                scaled y ticks=true,
                legend style={
                at={(0.5,1.3)},
                anchor=center,
                draw=black,
                legend cell align=left,
                fill=none,
                legend columns=2,
                font=\small
                }
                ]
                \addplot [mark=triangle*,color=spec_blue, thick] table [
                    x index=0, 
                    y index=1
                    ] {STLthO_DLW_noInsulation.txt};
                \addlegendentry{STL}
                \addplot [densely dotted,color=spec_orange, thick] table [
                    x index= 0, 
                    y index= 1 
                    ] {massImpedanceLaw_DLW.txt};
                \addlegendentry{Mass Impedance}
                \addplot [densely dotted,color=red, thick] table [
                    x index= 0, 
                    y index= 1 
                    ] {twelveDbOct_DLW_noInsulation.txt};
                \addlegendentry{12 dB/Octave}
                \addplot [dashed,color=spec_green, thick] coordinates {(84.92,0) (84.92,70)};
                \addlegendentry{Resonance Double-Leaf Wall}
            \end{semilogxaxis}
        \end{tikzpicture}
    \caption{Sound transmission loss (STL) of the large-scale building acoustics test facility with a double-leaf wall test specimen without insulation in one-third-octave bands over the frequency range $10\,\text{Hz}$ to $630\,\text{Hz}$. The resonance frequency of the double-leaf wall is $f_0=84.92\,\text{Hz}$.}
    \label{fig:STL_DLW_noInSu}
\end{figure}

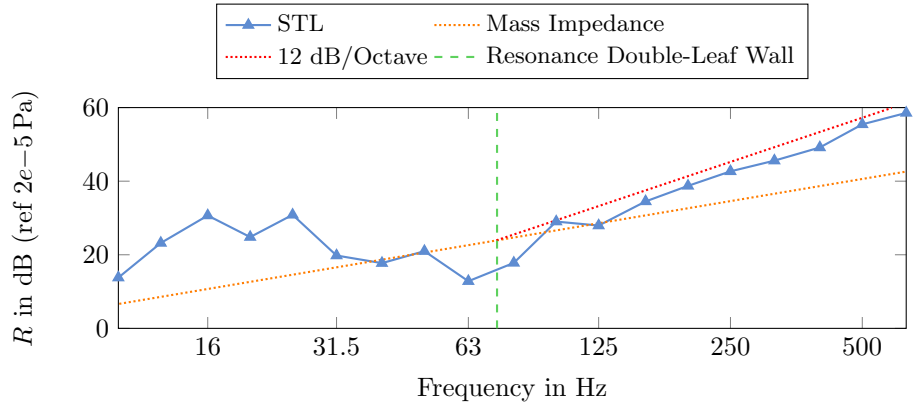
\begin{figure}[b!]
    \centering
        \begin{tikzpicture}
            \begin{semilogxaxis}[
                xlabel={Frequency in Hz},
                ylabel={$R$ in dB (ref $2e{-5}\,\text{Pa}$)},
                xmin=10, xmax=630,
                ymin=0, ymax=60,
                xtick={16,31.5,63,125,250,500},
                xticklabels={16,31.5,63,125,250,500},
                width=12cm,
                height=4.5cm,
                scaled y ticks=true,
                legend style={
                at={(0.5,1.3)},
                anchor=center,
                draw=black,
                legend cell align=left,
                fill=none,
                legend columns=2,
                font=\small
                }
                ]
                \addplot [mark=triangle*,color=spec_blue, thick] table [
                    x index=0, 
                    y index=1
                    ] {STLthO_DLW_Insulation.txt};
                \addlegendentry{STL}
                \addplot [densely dotted,color=spec_orange, thick] table [
                    x index= 0, 
                    y index= 1 
                    ] {massImpedanceLaw_DLW.txt};
                \addlegendentry{Mass Impedance}
                \addplot [densely dotted,color=red, thick] table [
                    x index= 0, 
                    y index= 1 
                    ] {twelveDbOct_DLW_Insulation.txt};
                \addlegendentry{12 dB/Octave}
                \addplot [dashed,color=spec_green, thick] coordinates {(73.26,0) (73.26,70)};
                \addlegendentry{Resonance Double-Leaf Wall}
            \end{semilogxaxis}
        \end{tikzpicture}
    \caption{Sound transmission loss (STL) of the large-scale building acoustics test facility with a double-leaf wall test specimen with insulation in one-third-octave bands over the frequency range $10\,\text{Hz}$ to $630\,\text{Hz}$. The resonance frequency of the double-leaf wall is $f_0=73.26\,\text{Hz}$.}
    \label{fig:STL_DLW_InSu}
\end{figure}

The theoretical profile of a double-leaf wall, according to the literature \cite{SIN2020}, \cite{VIG2008} or \cite{WIL2020}, is also divided into several areas. For small frequencies, the double leafs oscillating in phase which means that they behave as a single-leaf wall with the mass of both leafs and, thus, they obey Berger's mass impedance law in Equation (\ref{eqn:Berger}). Then, a decline occurs due to the resonance frequency of the double-leaf wall, where both leafs oscillate against each other. Following this, the STL increases with $12\,\text{dB}$ to $18\,\text{dB}$ per octave. However, more declines are existent due to the coincidence frequency (see Equation (\ref{eqn:Coincide})) of the leafs and cavity resonances in the gap. Those cavity resonance can be damped by placing an insulation into the gap. Since those cavity resonance lie outside the considered frequency range, they are not further introduced at this point. \\
The resonance frequency of the double leaf wall can be estimated by the resonance frequency of a two-mass-spring-oscillator \cite{VIG2008}
\begin{equation}
    f_0=\frac{1}{2\pi}\sqrt{\frac{s'}{m''_1}+\frac{s'}{m''_2}},    
\end{equation}
where $m_i''=\rho_i h_i$, $i=1,2$, is the areal mass of the $i$-th leaf of thickness $h_i$ and density $\rho_i$, and $s'=K/l_{x_\text{G}}$ is the dynamic stiffness. Here, $K$ equals $\text{Re}(\hat{K})$ if no insulation is applied and $\text{Re}(\hat{K}_e)$ if insulation is applied. It is to mention that this estimation formula does not incorporate any information regarding the support conditions of the leafs. Lastly, before discussing the numerically predicted STL profiles of the double-leaf wall test specimen, the coincidence frequency of a leaf of thickness $h=0.025\,\text{m}$ is $f_c=1324.65\,\text{Hz}$ and lies therefore outside the considered frequency range.\\
Figure \ref{fig:STL_DLW_noInSu} shows the STL of the double-leaf wall test specimen without insulation.
Also, Berger's mass impedance law is displayed as well as a straight line indicating the predicted $12\,\text{dB}$ incline from literature after the decline due to the double-leaf wall resonance at $f_0=84.92\,\text{Hz}$. Equivalently to the single-leaf wall, the low frequency range is determined by the modes of the single leafs, which are not displayed for reasons of conciseness. Then, an increase according to Berger's mass impedance law can be recorded from the $31.5\,\text{Hz}$-third-octave band on until a decline occurs in the area of the double-leaf wall resonance. After that, however, the predicted $12\,\text{dB}$-increase of the STL remains but a $6\,\text{dB}$ increase according to Berger's mass impedance law can be observed. To figure out, if a $12\,\text{dB}$-increase exists or not, the examined frequency range needs to be extended. So, at this point, the prediction from literature remain unobserved and the double-leaf wall's STL is not superior to those of the single-leaf wall. \\
Next to this, Figure \ref{fig:STL_DLW_InSu} shows the STL of the double-leaf wall test specimen with glass wool as insulation, as well as Berger's mass impedance law, the predicted $12\,\text{dB}$-STL-incline, and the double-leaf wall resonance $f_0=73.26\,\text{Hz}$. 
Comparing the STL profile to those of the double-leaf wall without insulation in Figure \ref{fig:STL_DLW_noInSu}, the profiles are similar up to the third-octave band with the mid-frequency $40\,\text{Hz}$. As the double-wall resonance is smaller compared to the test specimen without insulation, the STL-decline occurs earlier, namely in the area of the estimated double-leaf wall resonance frequency. After that, the predicted $12\,\text{dB}$-incline per octave can be clearly observed.

\section{Conclusion}
This contribution presented a comprehensive overview of the modelling process to obtain a building acoustics test facility FE model according to DIN EN ISO 10140-5 which can be used to predict the STL of a single- and double-leaf wall with and without insulation utilising measurement regulations defined in DIN EN ISO 10140-4 and DIN EN ISO 10140-5. In doing so, the FE model was built using a simulation chain exploiting SALOME 9.14 as a geometry and mesh creator and the in-house research code elPaSo for matrix assembly and solving. Before predicting the STL of the large-scale building acoustics test facility numerically, elPaSo was verified for a small-scale test facility with a single- and double leaf wall with and without insulation using COMSOL 6.3. Subsequently, the numerical STL estimation in the large-scale building acoustics test facility was executed exclusively using SALOME and elPaSo without comparing the results to COMSOL. The reason was that it is not straightforward to couple two- and three-dimensional elements at an interface in COMSOL 6.3, which prevented the use of non-conforming meshes. 
Instead, COMSOL would have yield a resource- and time-exhaustive computation since the smallest element length, determined by the equivalent fluid, would have resulted in very fine discretised conforming FE meshes. To increase the efficiency of the numerical STL estimation, the simulation chain consisting of SALOME and elPaSo exploited the use of non-conforming FE meshes. However, at first, a mesh convergence study was undertaken for the large-scale building acoustics test facility. Then, in accordance with the results of the convergence study, the STL of the single- and double-leaf wall without and with insulation were computed in the frequency range $8\,\text{Hz}$ to $630\,\text{Hz}$ using an domain- and frequency-specific discretisation approach. Also, a comparison to theoretical STL profiles predicted in literature was performed. The double-leaf wall with insulation revealed a good agreement with theoretical STL profiles from literature, while the single-leaf and double-leaf wall without insulation only showed partial agreement. More justified statements regarding the comparison between the numerically estimated STLs and theoretical STL profiles require investigations in a wider frequency range. \\
The published benchmark of the simple building acoustic test facility FE model can be computed with a decent computational effort. However, this does not mark the limits of elPaSo but lays the foundation for future investigations that, first, extend the examined frequency range, and second focus on virtual prototyping of entire buildings in the sense of an acoustic-oriented design process. In doing so, efficiency-enhancing measures shall be exploited, for instance, the use of efficient direct solvers, surrogate modelling methods, e.g. MOR, and HPC clusters.\\
\\
\\
\textbf{Funding}\\
The authors received funding from the German Research Foundation (DFG) under the project "Efficient broad-band simulation of large-scale vibroacoustic systems with random input data" (project no. 531569940).\\
\\
\textbf{Data availability statement}\\
The dataset to compute the sound transmission loss of the building acoustics test facility with a single- and double-leaf wall with and without insulation is published on Zenodo hosted by the community EAA Computational Acoustics, under the reference \cite{SCH2025}.

%%%%%%%%%%%%%%%%%%%%%%%% numbered reference %%%%%%%%%%%%%%%%%%%%%%%%%%%%

%%%%%%%%%%%%%%%%%%%%%%%% numbered reference %%%%%%%%%%%%%%%%%%%%%%%%%%%%
\newpage
\FloatBarrier
\appendix
\section{elPaSo-Verification based on a Small-Scale Test Facility - Results}\label{App:VerifEl}
\begin{figure}[!h]
    \centering
        \begin{tikzpicture}
            \begin{axis}[
                xlabel={Frequency in Hz},
                ylabel={SPL in dB (ref $2e{-5}\,\text{Pa}$)},
                xmin=1, xmax=1000,
                width=12cm,
                height=4.5cm,
                scaled y ticks=true,
                legend style={
                at={(0.5,1.2)},
                anchor=center,
                draw=black,
                legend cell align=left,
                fill=none,
                legend columns=3,
                font=\small
                }
                ]
                \addplot [mark=none, color=spec_blue, thick] table [
                    x index= 0, 
                    y index= 1 
                    ] {meanSolution_elPaSo_AirTrev_77k_SR2.txt};
                \addlegendentry{elPaSo}
                \addplot [mark=none, color=orange, thick, dashed] table [
                    x index=0, 
                    y index=1
                    ] {meanSolution_soundPressure_SR2_1_1000Hz_dampedTrev.txt};
                \addlegendentry{COMSOL}
            \end{axis}
        \end{tikzpicture}
    \caption{Single-leaf wall configuration (SLW2): sound pressure level (SPL) at microphone position $\text{SR}_2$ computed using elPaSo and COMSOL, with an maximum relative error $\epsilon_{r,max}=0.04$ and average relative error $\epsilon_{r,ave}=0.00$.}
    \label{fig:SLW_dampTrev_SR2}
\end{figure}
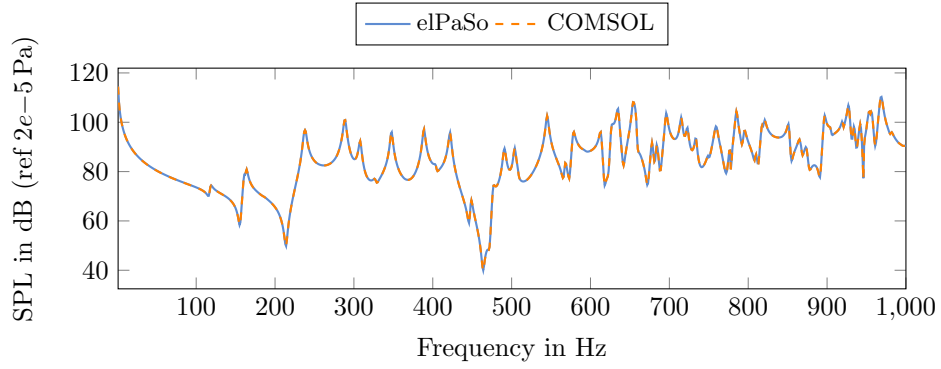
\vspace{3cm}
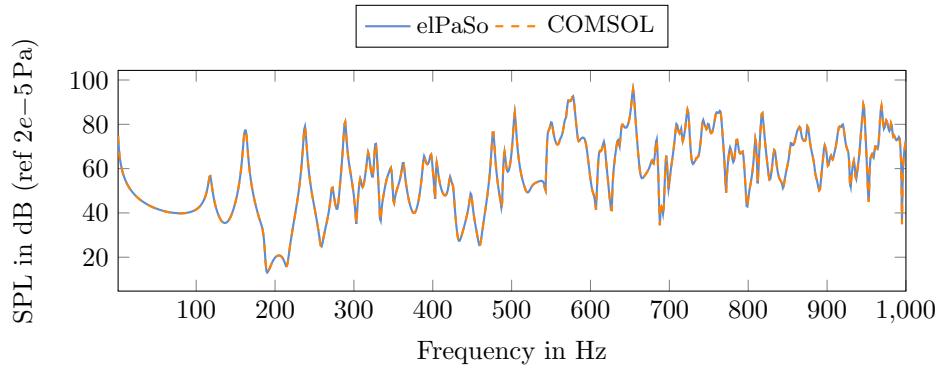
\begin{figure}[!h]
    \centering
        \begin{tikzpicture}
            \begin{axis}[
                xlabel={Frequency in Hz},
                ylabel={SPL in dB (ref $2e{-5}\,\text{Pa}$)},
                xmin=1, xmax=1000,
                width=12cm,
                height=4.5cm,
                scaled y ticks=true,
                legend style={
                at={(0.5,1.2)},
                anchor=center,
                draw=black,
                legend cell align=left,
                fill=none,
                legend columns=3,
                font=\small
                }
                ]
                \addplot [mark=none, color=spec_blue, thick] table [
                    x index= 0, 
                    y index= 1 
                    ] {meanSolution_elPaSo_AirTrev_77k_RR2.txt};
                \addlegendentry{elPaSo}
                \addplot [mark=none, color=orange, thick, dashed] table [
                    x index=0, 
                    y index=1
                    ] {meanSolution_soundPressure_RR2_1_1000Hz_dampedTrev.txt};
                \addlegendentry{COMSOL}
            \end{axis}
        \end{tikzpicture}
    \caption{Single-leaf wall configuration (SLW2): sound pressure level (SPL) at microphone position $\text{RR}_2$ computed using elPaSo and COMSOL, with an maximum relative error $\epsilon_{r,max}=0.18$ and average relative error $\epsilon_{r,ave}=0.01$.}
    \label{fig:SLW_dampTrev_RR2}
\end{figure}
\begin{figure}[!h]
    \centering
        \begin{tikzpicture}
            \begin{axis}[
                xlabel={Frequency in Hz},
                ylabel={SPL in dB (ref $2e{-5}\,\text{Pa}$)},
                xmin=1, xmax=1000,
                width=12cm,
                height=4.5cm,
                scaled y ticks=true,
                legend style={
                at={(0.5,1.2)},
                anchor=center,
                draw=black,
                legend cell align=left,
                fill=none,
                legend columns=3,
                font=\small
                }
                ]
                \addplot [mark=none, color=spec_blue, thick] table [
                    x index= 0, 
                    y index= 1 
                    ] {meanSolution_elPaSo_Insulation_337k_SR2.txt};
                \addlegendentry{elPaSo}
                \addplot [mark=none, color=orange, thick, dashed] table [
                    x index=0, 
                    y index=1
                    ] {meanSolution_soundPressure_SR2_1_1000Hz_Insulation.txt};
                \addlegendentry{COMSOL}
            \end{axis}
        \end{tikzpicture}
    \caption{Double-leaf wall configuration with insulation (DLWI): sound pressure level (SPL) at microphone position $\text{SR}_2$ computed using elPaSo and COMSOL, with an maximum relative error $\epsilon_{r,max}=0.01$ and average relative error $\epsilon_{r,ave}=0.0$.}
    \label{fig:DLW_Insu_SR2}
\end{figure}
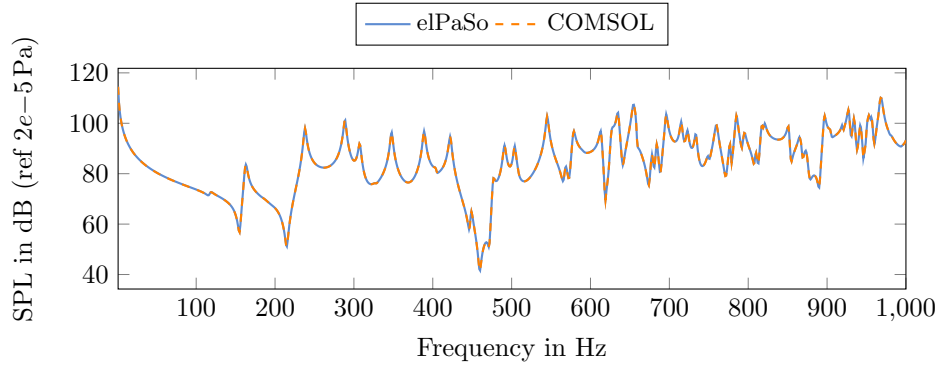
\vspace{2.5cm}
\begin{figure}[!h]
    \centering
        \begin{tikzpicture}
            \begin{axis}[
                xlabel={Frequency in Hz},
                ylabel={SPL in dB (ref $2e{-5}\,\text{Pa}$)},
                xmin=1, xmax=1000,
                width=12cm,
                height=4.5cm,
                scaled y ticks=true,
                legend style={
                at={(0.5,1.2)},
                anchor=center,
                draw=black,
                legend cell align=left,
                fill=none,
                legend columns=3,
                font=\small
                }
                ]
                \addplot [mark=none, color=spec_blue, thick] table [
                    x index= 0, 
                    y index= 1 
                    ] {meanSolution_elPaSo_Insulation_337k_RR2.txt};
                \addlegendentry{elPaSo}
                \addplot [mark=none, color=orange, thick, dashed] table [
                    x index=0, 
                    y index=1
                    ] {meanSolution_soundPressure_RR2_1_1000Hz_Insulation.txt};
                \addlegendentry{COMSOL}
            \end{axis}
        \end{tikzpicture}
    \caption{Double-leaf wall configuration with insulation (DLWI): sound pressure level (SPL) at microphone position $\text{RR}_2$ computed using elPaSo and COMSOL, with an maximum relative error $\epsilon_{r,max}=0.03$ and average relative error $\epsilon_{r,ave}=0.01$.}
    \label{fig:DLW_Insu_RR2}
\end{figure}
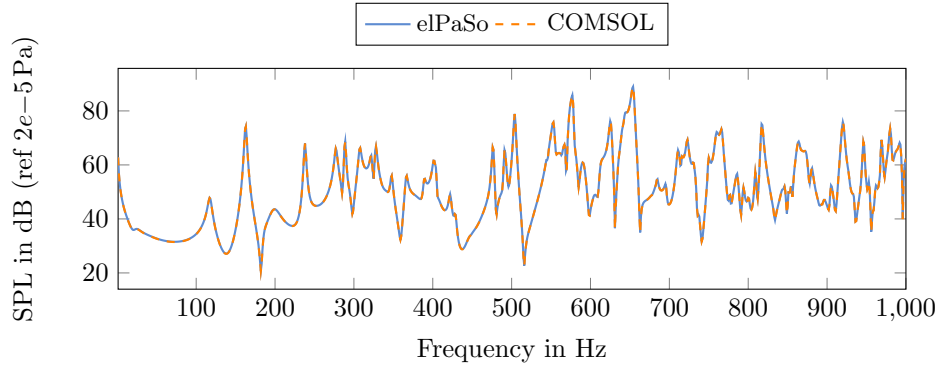

\clearpage
%\FloatBarrier
\section{Large-Scale Test Facility - Mesh Convergence Study}\label{App:BATF_Res}
\FloatBarrier
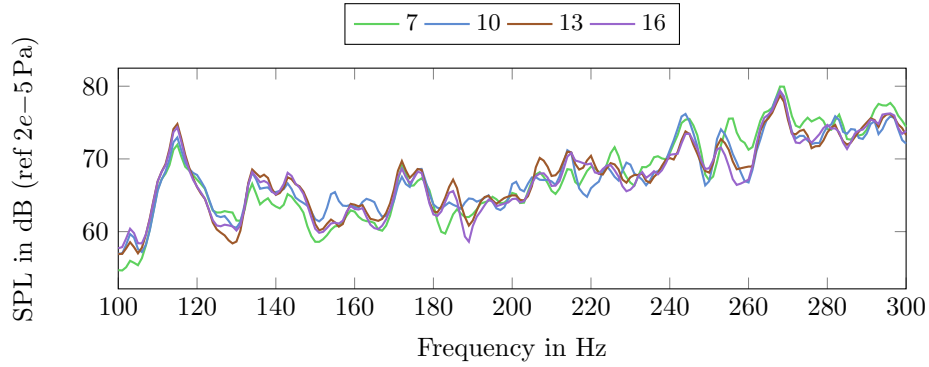
\begin{figure}[!h]
    \centering
        \begin{tikzpicture}
            \begin{axis}[
                ylabel={SPL in dB (ref $2e{-5}\,\text{Pa}$)},
                xlabel={Frequency in Hz},
                xmin=100, xmax=300,
                width=12cm,
                height=4.5cm,
                scaled y ticks=true,
                legend style={
                at={(0.5,1.2)},
                anchor=center,
                draw=black,
                legend cell align=left,
                fill=none,
                legend columns=4,
                font=\small
                }
                ]
                \addplot [color=spec_green, thick] table [
                    x index=0, 
                    y index=1
                    ] {ConvStud_7nodes_Insulation_SR.txt};
                \addlegendentry{7}
                \addplot [color=spec_blue, thick] table [
                    x index= 0, 
                    y index= 1 
                    ] {ConvStud_10nodes_Insulation_SR.txt};
                \addlegendentry{10}
                \addplot [color=spec_brown, thick] table [
                    x index= 0, 
                    y index= 1 
                    ] {ConvStud_13nodes_Insulation_SR.txt};
                \addlegendentry{13}
                \addplot [color=spec_amethyst, thick] table [
                    x index= 0, 
                    y index= 1 
                    ] {ConvStud_16nodes_Insulation_SR.txt};
                \addlegendentry{16}
            \end{axis}
        \end{tikzpicture}
    \caption{Mean sound pressure level (SPL) over all receiver positions in the source room of the large-scale building acoustics test facility in the frequency range $100\,\text{Hz}$ to $300\,\text{Hz}$ based on meshes with quadratic elements utilising element lengths according to 7, 10, 13 and 16 nodes per wavelength.}
    \label{fig:SPL_7_16nodes_SR}
\end{figure}
\vspace{3cm}
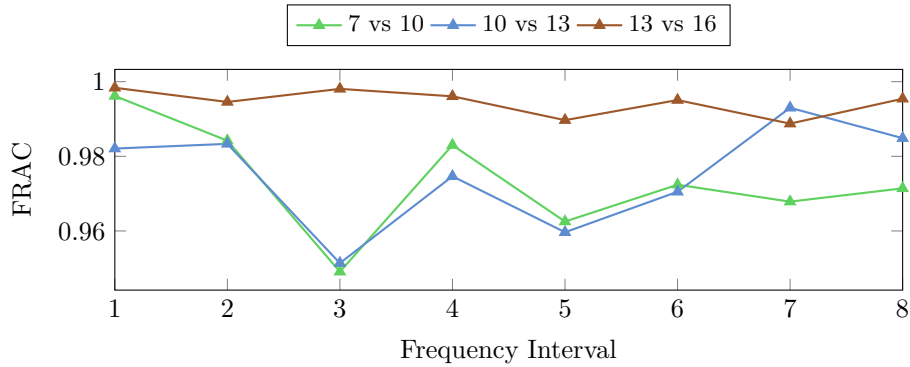
\begin{figure}[!h]
    \centering
        \begin{tikzpicture}
            \begin{axis}[
                ylabel={FRAC},
                xlabel={Frequency Interval},
                xmin=1, xmax=8,
                width=12cm,
                height=4.5cm,
                scaled y ticks=true,
                legend style={
                at={(0.5,1.2)},
                anchor=center,
                draw=black,
                legend cell align=left,
                fill=none,
                legend columns=3,
                font=\small
                }
                ]
                \addplot [mark=triangle*, color=spec_green, thick] table [
                    x index=0, 
                    y index=1
                    ] {Frac_7_10nodes_Insulation_SR.txt};
                \addlegendentry{7 vs 10}
                \addplot [mark=triangle*, color=spec_blue, thick] table [
                    x index= 0, 
                    y index= 1 
                    ] {Frac_10_13nodes_Insulation_SR.txt};
                \addlegendentry{10 vs 13}
                \addplot [mark=triangle*, color=spec_brown, thick] table [
                    x index= 0, 
                    y index= 1 
                    ] {Frac_13_16nodes_Insulation_SR.txt};
                \addlegendentry{13 vs 16}
            \end{axis}
        \end{tikzpicture}
    \caption{Mesh convergence study of the large scale building acoustics test facility with a double-leaf wall and insulation utilising the FRAC criterion in $25\,\text{Hz}$ intervals in the frequency range $100\,\text{Hz}$ to $300\,\text{Hz}$. Here, the mean sound pressure at all receiver positions in the source room are compared arising from a mesh with quadratic elements where the element length is determined by samplings with 7, 10, 13 and 16 nodes per wavelength.}
    \label{fig:FRAC_SR}
\end{figure}

\end{document}